\documentclass[useAMS,usenatbib,referee]{mn2e}
\usepackage{supertabular}
 \title[UKIDSS-2MASS Proper Motion Survey I]{The UKIDSS-2MASS Proper Motion Survey I: Ultracool dwarfs from UKIDSS DR4}
 \author[N.R.\ Deacon et al.]{N.R.\ Deacon\thanks{E-mail:
 nd@roe.ac.uk}$^1$ N.C.\ Hambly$^2$, R.R. King$^3$ and M.J. McCaughrean$^3$\\
$^1$Department of Astrophysics, Faculty of Science, Radboud University Nijmegen, P.O. Box 9010, 6500 GL Nijmegen, The Netherlands\\
$^2$SUPA\thanks{Scottish Universities' Physics Alliance}, Institute for Astronomy, 
School of Physics, University of Edinburgh, Royal Observatory Edinburgh, Blackford Hill,\\
Edinburgh, EH9 3HJ, UK\\
$^3$Astrophysics Group, School of Physics, University of Exeter, Stocker Road, Exeter, EX4 4QL, UK}
 \begin{document}
 \date{}
 \pagerange{\pageref{firstpage}--\pageref{lastpage}} \pubyear{2005}
 \maketitle
 \label{firstpage}
 \begin{abstract}
The UKIRT Infrared Deep Sky Survey (UKIDSS) is the first of a new generation of infrared surveys. Here we combine the data from two UKIDSS components, the Large Area Survey (LAS) and the Galactic Cluster Survey (GCS), with 2MASS data to produce an infrared proper motion survey for low mass stars and brown dwarfs. In total we detect 267 low mass stars and brown dwarfs with significant proper motions. We recover all ten known single L dwarfs and the one known T dwarf above the 2MASS detection limit in our LAS survey area and identify eight additional new candidate L dwarfs. We also find one new candidate L dwarf in our GCS sample. Our sample also contains objects from eleven potential common proper motion binaries. Finally we test our proper motions and find that while the LAS objects have proper motions consistent with absolute proper motions, the GCS stars may have proper motions which are significantly under-estimated. This is due possibly to the bulk motion of some of the local astrometric reference stars used in the proper motion determination.
 \end{abstract}
 \section{Introduction} 
The study of the lowest mass stars and brown dwarfs is one of the most active areas of current Galactic research. The identification of samples of such objects allows both the low mass/substellar luminosity and mass functions to be constrained. These can in turn be used to constain models of star and brown dwarf formation. Additionally such samples may produce interesting single or multiple objects, whose spectroscopic properties could inform atmospheric models for brown dwarfs and giant planets.

Until relatively recently the majority of proper motion surveys have concentrated on optical data. Among the first to use proper motion as a tool to select faint, nearby populations was Luyten. His work using photographic plates culminated in two large scale catalogues of high proper motion stars, the Luyten Half Arcsecond catalogue (LHS - Luyten 1979a) and the New Luyten Two-Tenths catalogue (NLTT - Luyten 1979b). Since that time much work has gone into identifying objects Luyten missed and filling in areas of poor coverage. The majority of recent wide-field proper motion surveys have used digitised photographic plate data, mostly in optical filters. Such surveys include Lepine \& Shara (2008), Hambly et al. (2004), Pokorny et al. (2003) and Finch et al. (2007). Infrared proper motion surveys are an ideal instrument for producing samples of low mass stars and brown dwarfs. These objects are cool $T_{eff}<3500K$ and hence are intrinsically faint and red. As a result they are most easily detected in the infrared. The first infrared proper motion survey was conducted by Deacon, Hambly \& Cooke (2005), this used SuperCOSMOS (Hambly et al. 2001) scans of UKST $I$ plates along with data from the 2 Micron All Sky Survey (2MASS - Skrutskie et al., 2006) to produce a sample of 144 low mass stars and brown dwarfs with proper motions above half an arcsecond per year. In the later paper (Deacon \& Hambly 2007) the sample was expanded to more than 7000 objects with the minimum proper motion limit reduced to 0.1''/yr. Recently Looper et al. (2008) used available overlap areas in the 2MASS data to detect two new L dwarfs using an infrared proper motion survey. Metchev et al. (2008) used 2MASS data along with those from the partially infrared Sloan Digital Sky Survey (Adelman-McCarthy et al. 2006) to identify two new T dwarfs and 22 new L dwarfs.

Recent large-scale infrared surveys such as the DEep Near Infrared Survey (DENIS - Epchtein et al., 1997)) and 2MASS took place at the end of the last decade and the start of this one. They produced CCD quality infrared data over tens of thousands of square degrees to depths of 16.5 and 15.8 respectively in the $J$ band. Both surveys have led to large samples of late type objects being identified and classified (Kirkpatrick et al. 1999, Burgasser et al. 2002, Delfosse et al. 1997). The generation of infrared surveys following DENIS and 2MASS is led by the UKIRT Infrared Deep Sky Survey (UKIDSS - Lawrence et al. 2007). This suite of surveys using WFCAM (Casali et al. 2007) on the UK Infrared Telescope (UKIRT) provides both large scale surveys for studies of galactic and extragalactic populations and deep pencil-beam surveys to examine the population of high redshift galaxies. The two surveys of the most interest to the study of low mass stars and brown dwarfs are the Large Area Survey (LAS) and the Galactic Clusters Survey (GCS). The LAS will have a final area of 4000 square degrees, will be in four passbands ($Y$, $J$, $H$ and $K$) to a depth of $J$=19.6 and will also provide a second $J$ band epoch for proper motion measurements. It covers an area away from the Galactic plane and is intended for the study of high redshift quasars and cool dwarf/subdwarfs in the field. This survey has already produced interesting results, both in terms of individual cool objects (Warren et al. 2007) and the population of low mass objects (Pinfield et al. 2008). The GCS concentrates on ten star forming regions and open clusters. In addition to the four colours used in the LAS, $Z$ band data is available and a second $K$ epoch will provide proper motion data for the clusters. So far studies such as Lodieu et al. (2006) and Lodieu et al. (2007) have used the data released to produce mass functions for Upper Sco and the Pleiades respectively. The GCS can be used also for the study of objects lying in the foreground of the target clusters (Lodieu et al., in preparation). The UKIDSS data is released through the WFCAM Science Archive (WSA - Hambly et al. 2008) and is available to all European Southern Observatories (ESO) members. The current release is Data Release 4 (DR4). Additionally astronomers from non-ESO countries can access the data after an 18 month delay. The soon to be operational Visual and Infrared Survey Telescope for Astronomy (VISTA - Emerson et al. 2004) will provide an even faster infrared surveying capability than WFCAM. It will cover the whole southern hemisphere in two bands ($J$ and $K$) including several thousand square degrees with additional $Y$ and/or $H$ band photometry.
 \section{Method} 
 \begin{figure}
 \setlength{\unitlength}{1mm}
 \begin{picture}(75,120)
 \includegraphics{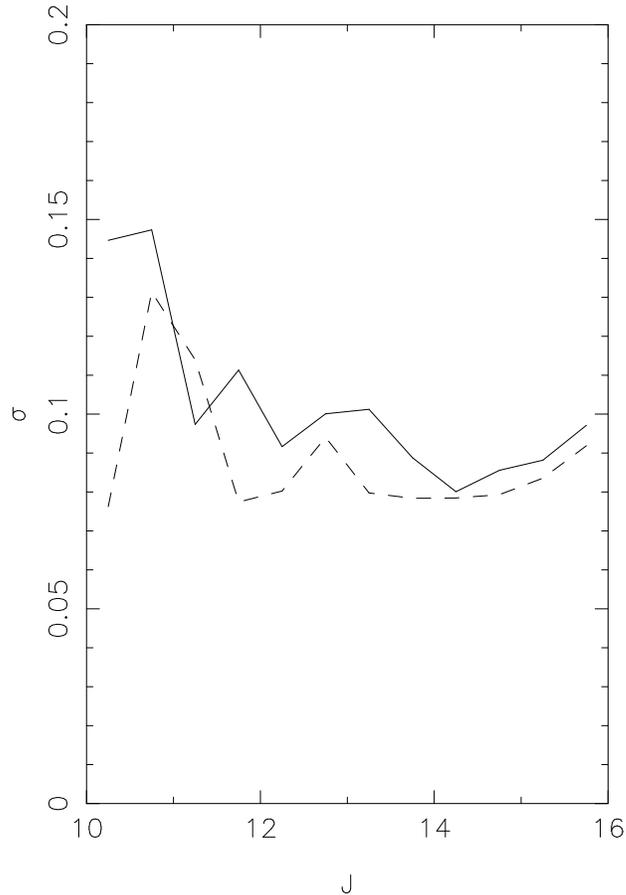}
 \end{picture}
 \caption[]{The astrometric errors between the UKIDSS and 2MASS data. The solid line represents the error in Right Ascension and the dashed line the error in Declination.}
 \label{Errplot}
 \end{figure}
Our method for selecting candidate ultracool dwarfs is almost identical to that used in Deacon et al. (2005) and Deacon \& Hambly (2007). The first selection process is to identify UKIDSS objects with counterparts in 2MASS within a distance of roughly 4-5$\sigma$ (where $\sigma$ is the positional error between the surveys). Clearly to be able to carry out this selection we must first identify the typical positional errors between the surveys. This was done by comparing the positions of objects between the two surveys and calculating the standard deviation between the two while removing non-Gaussian outliers. A plot of the errors vs. $J$ band magnitude is shown in Figure~\ref{Errplot}. It is reasonable to assume that above the saturation limit ($J$=10.5) the positional errors are of the order of 0.1 arcseconds in each axis. Hence a minimum positional shift of 0.6 arcseconds will have a significance of $\ge5\sigma$.

In addition to the astrometric cut, only objects with good stellar profiles (i.e. a classification statistic between $-3$ and 3) were selected. We also excluded objects with magnitudes brighter than the saturation limit for each filter and those more than one magnitude fainter than the 2MASS detection limit of $J$=15.8. Next we  examined the ($Y-J$) vs. ($J-H$) colour-colour plot found in Leggett et al. (2005) and set a cut that either ($Y-J$) had to be greater than 0.7 or ($Y-J$) had to be greater than 0.3 and ($J-H$) less than zero. This allowed us to exclude the main mass of main sequence stars while still covering the areas where L and T dwarfs (and the Baraffe et al. 2003 models for Y dwarfs) are expected to lie. All of these cuts were enacted using the SQL query shown in Appendix A. This was submitted to the WFCAM Science Archive which returned our unpaired UKIDSS sample.

Next we attempted to identify 2MASS companions to the unpaired UKIDSS objects. This was done by searching through the appropriate 2MASS files and identifying any good stellar object within 100 arcseconds of the UKIDSS object with a similar J band magnitude (within one magnitude). To ensure that each 2MASS object was not a non-moving coincidence background object, we then checked to ensure each object did not have a UKIDSS companion within the 0.6 arcsecond movement cut. This was done using the WSA CrossID function. Any 2MASS object with a UKIDSS pair within 0.6 arcseconds was excluded from the sample. Next the UKIDSS images of all the candidates were inspected by eye to remove any objects which could have poor astrometry resulting from deblended images or nearby bright stars.

To ensure accurate proper motion estimates we attempted to calculate individual astrometric solutions for each object in our sample. To do this we used the WSA Cross-ID function to identify all good stellar images within five arcminutes of each object in our sample. These objects provide a set of astrometric reference stars for each star in our sample. We then cross referenced these reference stars with the 2MASS catalogue so we had their positions in both datasets. For each star in our sample we selected reference stars with J band brightnesses within one magnitude of the target and used these to do a standard six parameter plate-to-plate fit. This allows us to correct for any offsets between the astrometric systems and calculate local positional errors. If a target had too few reference stars (five or fewer) associated with it, no plate-to-plate fit was carried out and a magnitude dependent error estimate was taken from the global error estimate shown in Figure~\ref{Errplot}. Whether an object's astrometry uses the local or global astrometric solutions is indicated in the data tables (see Appendix B). Only objects with proper motions more significant than 5$\sigma$ were included in the final sample. However an astrometric solution calculated using too few reference stars will underestimate the positional errors. Hence we carried out a series of simulations to estimate difference between the positional errors found using a set of $n_{ref}$ ($\sigma_{measured}$) and the true positional error ($\sigma_{true}$) which we set as an input to our simulations. It was found that the following relationship was a good estimate for the underestimation of the positional errors, 
\begin{equation}
\frac{\sigma_{true}}{\sigma_{measured}} \approx 1 + \frac{19.8}{n_{ref}^{1.5}}
\end{equation}
 Hence this equation was used to apply a correction factor to all the astrometric solutions calculated using sets of reference stars. The main aim of this study is to examine the population of ultracool dwarfs in the field. This is complicated by the fact that in the GCS areas, our 5$\sigma$ lower proper motion cut could include some objects which are members of the target clusters. Hence for the GCS data we have set a minimum proper motion of 80 milliarcseconds per year. This excludes objects with proper motions similar to all the GCS clusters with the exception of the Hyades (which does not appear in the dataset we use). 

Our survey covers all the area included in the UKIDSS DR4 for the LAS and the GCS. These surveys consist of 993 sq.deg. and 148 sq.deg. of sky respectively.
 \section{Results}
 \begin{figure}
 \setlength{\unitlength}{1mm}
 \begin{picture}(75,180)
 \includegraphics{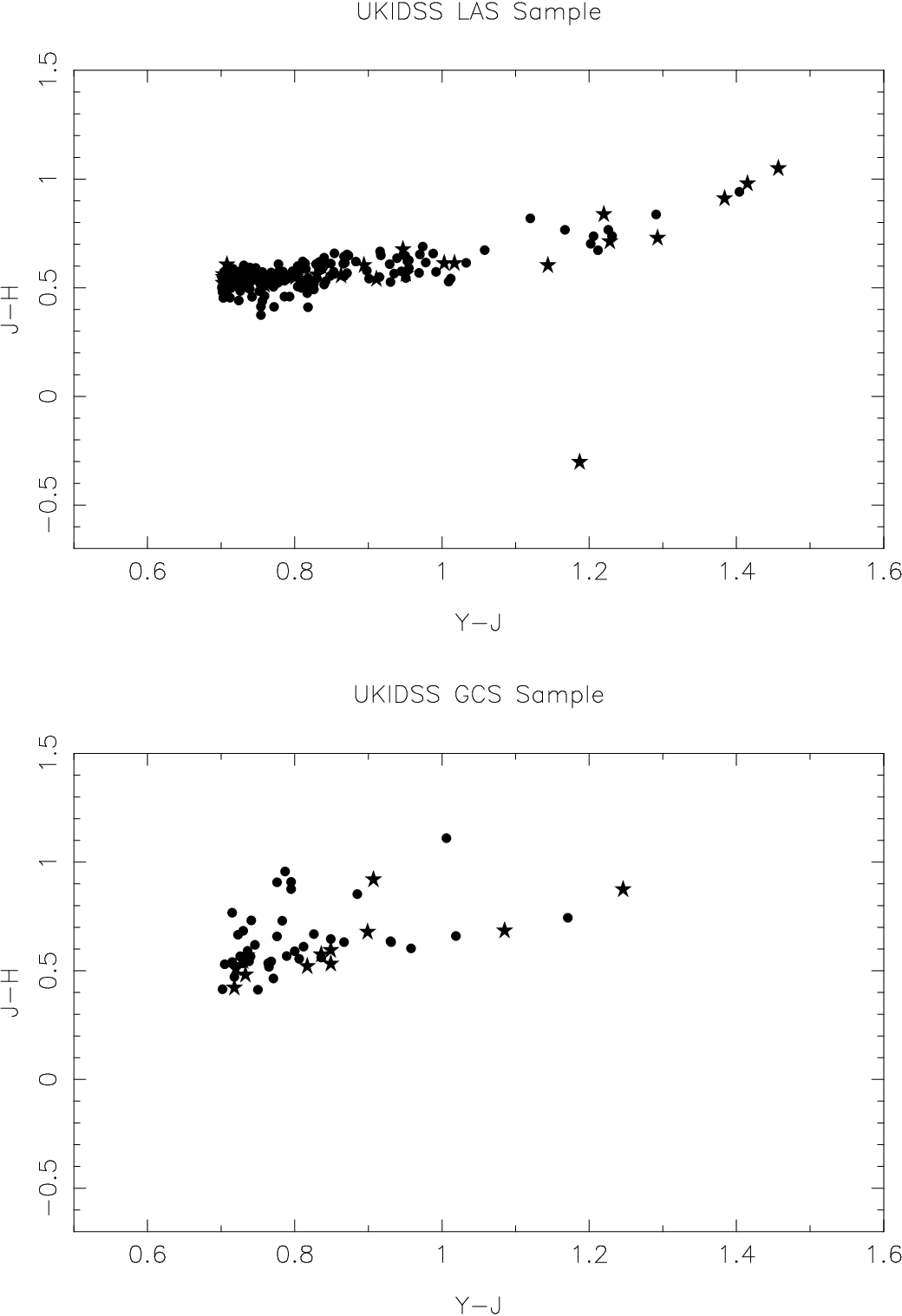}
 \end{picture}
 \caption[]{The UKIDSS colour-colour diagrams for our two samples. Dots represent newly discovered objects and stars those found in previous surveys. Note the second locus in the GCS data which lies above the main locus. We believe this consists of highly reddened early type stars behind molecular clouds.}
 \label{ccdiagram}
 \end{figure}
The Large Area Survey (LAS) sample produced 213 candidate objects of which 44 are previously identified by other surveys. The Galactic Clusters Survey (GCS) produced a sample of 54 objects, 12 of which are previously known. A list of all the objects identified can be found in Appendix B. Figure~\ref{ccdiagram} shows infrared colour--colour diagrams for both survey samples. In both cases it is clear that there is only one object which is both redder than $Y-J$=1.1 and bluer than $J-H$=0.5. Comparing these with colours from Hewett et al. (2006) we can conclude that there is only one object with a spectral type later than T3 in our sample (our full sample includes all objects found in the LAS and GCS). One clear difference between the two samples of objects found in the LAS and those found in the GCS is the appearance of several objects above the main M dwarf locus in the GCS sample. By comparison the LAS sample has a tight stellar locus which becomes less dense towards the L dwarf regime (redder than $Y-J$=1.1). One member of this second locus is NLTT 42735, an object which appears in the initial DENIS low mass stars sample of Crifo et al. (2005). They use spectroscopy to identify NLTT 42735 (UGCS2MASS1625$-$2400) as a reddened F--K type star and hence exclude it from their final sample. They point out that its reddening is probably due to its proximity to a known molecular cloud. Checking our seven objects (UGCS2MASS0408+2447, UGCS2MASS0536-0454, UGCS2MASS0433+2933, UGCS2MASS0541--0305, UGCS2MASS1625--2400, UGCS2MASS1631--2404 and UGCS2MASS1637-2200) which lie above the main stellar locus using the SIMBAD database reveals that five are within ten arcminutes of know molecular clouds or dark nebulae. Given that this second locus only appears in the GCS (which samples clusters which lie close to star formation regions) and not in the LAS (which samples an area of high Galactic latitude) we believe we can assume that it consists of earlier--type stars which lie behind the star formation regions and are reddened by the gas and dust. As we do not believe these objects are true low mass stars or brown dwarfs, we recommend that they are excluded from any cool dwarf sample derived from our data. As a number of other objects appear to lie only slightly above the main stellar locus we would advise anyone using our GCS sample to bear in mind that these objects may be reddened early type stars also.

\subsection{Ultracool Dwarfs}  
The prime motivator for this survey is to identify a clean sample ultracool dwarfs. To further this aim we selected objects in our sample which were redder than $Y-J$=1.1. According to Hewett et al. (2006), this should give us a sample completely free from M dwarfs and earlier spectral types. The objects identified are listed in Table~\ref{Ldwarfs} along with their astrometry and photometry. Of the total of nineteen objects listed, eleven are previously known (one of which is a previously identified T dwarf). Hence of the eighteen objects with L dwarf--like colours only ten are previously known. Clearly spectroscopic identification of the previously unidentified eight objects is required, but even if only a few are genuine L dwarfs, this would indicate incompleteness in our current knowledge of even the relatively bright L dwarf population.
\begin{table*}
 \begin{minipage}{160mm}
  \caption{A list of all objects in our sample redder than $Y-J$=1.1. The astrometric solutions were calculated using a fit to local reference stars. All photometry uses the standard UKIDSS filters (Hewett et al. 2006). Citation key --- a:Schneider et al. (2002), b: Fan et al. (2000), c: Bouy et al. (2003), d: Reid et al. (2008), e: Burgasser et al. (2004), f: Wilson et al. (2003), g: Kirkpatrick et al. (2000), h: Knapp et al. (2004), i: Hawley et al. (2002).}
\label{Ldwarfs}
\tiny
  \begin{tabular}{lccrcrccccl}
  \hline
Name&Position&$\mu$&P.A&$\sigma_{\mu}$&$\sigma_{PA}$&$Y$&$J$&$H$&$K$&note\\
&&''/yr&$^{\circ}$&''/yr&&&&&&\\
  \hline
ULAS2MASS0001+1535&00 01 12.24 +15 35 34.3&0.218&144.0&0.029$^1$&8.4&16.877&15.462&14.483&13.623&h\\ 
ULAS2MASS0054-0031&00 54 06.66 $-$00 31 03.2&0.248&129.8&0.029$^1$&6.9&16.857&15.629&14.917&14.302&a\\
ULAS2MASS0205+1251&02 05 03.66 +12 51 42.0&0.362&93.8&0.025$^1$&4.3&17.018&15.561&14.513&13.637&g\\ 
ULAS2MASS0207+1355&02 07 35.73 +13 55 54.9&0.316&123.3&0.017$^1$&3.6&16.589&15.369&14.532&13.829&i\\ 
ULAS2MASS0219+0506&02 19 22.05 +05 06 30.8&0.184&84.1&0.013$^1$&5.5&16.022&14.855&14.089&13.453&\\ 
ULAS2MASS0330-0025&03 30 35.32 $-$00 25 37.4&0.509&131.1&0.015$^1$&1.7&16.508&15.215&14.487&13.771&b\\ 
ULAS2MASS0344+0111&03 44 08.88 +01 11 23.9&0.209&211.1&0.030$^1$&6.1&15.698&14.554&13.952&13.446&c\\
UGCS2MASS0409+2104&04 09 09.57 +21 04 37.9&0.184&150.6&0.007$^1$&2.4&16.622&15.376&14.502&13.781&i\\
ULAS2MASS0835+0548&08 35 58.24 +05 48 30.7&0.110&257.6&0.009$^1$&7.7&15.705&14.499&13.762&13.127&d\\ 
ULAS2MASS0843+1024&08 43 33.31 +10 24 43.0&0.593&165.3&0.011$^1$&1.5&15.989&14.777&14.105&13.551&\\ 
ULAS2MASS1211+0406&12 11 30.11 +04 06 08.1&0.225&195.7&0.035$^1$&6.5&16.808&15.517&14.680&13.949&\\
ULAS2MASS1231+0847&12 31 46.99 +08 47 25.8&1.590&228.1&0.023$^1$&0.9&16.340&15.153&15.456&15.552&e\\ 
ULAS2MASS1308+0818&13 08 30.97 +08 18 52.5&0.233&281.1&0.022$^1$&3.4&16.312&15.192&14.373&13.792&\\
ULAS2MASS1346+0842&13 46 07.37 +08 42 34.1&0.235&246.6&0.042$^1$&8.9&16.744&15.518&14.752&14.115&\\ 
ULAS2MASS1407+1241&14 07 53.45 +12 41 10.4&0.337&280.5&0.022$^1$&3.3&16.737&15.333&14.392&13.632&d\\ 
ULAS2MASS1422+0827&14 22 57.10 +08 27 50.4&0.592&195.6&0.012$^1$&1.5&16.240&15.009&14.271&13.609&\\ 
ULAS2MASS1448+1031&14 48 25.75 +10 31 58.1&0.278&117.5&0.018$^1$&5.1&15.804&14.420&13.510&12.674&f\\
ULAS2MASS1452+1114&14 52 01.96 +11 14 56.9&0.394&139.4&0.027$^1$&4.1&16.771&15.569&14.867&14.280&\\ 
UGCS2MASS1630-2120&16 30 17.69 $-$21 20 01.4&0.153&259.8&0.013$^1$&3.8&15.695&14.524&13.780&13.174&\\
  \hline
\normalsize
\end{tabular}
\end{minipage}
\end{table*}
\subsection{Common Proper Motion Objects} 
\begin{table*}
 \begin{minipage}{160mm}
  \caption{Photometry and astrometry for the two objects ULAS2MASS1253+0740a and ULAS2MASS1253+0740b which share a common proper motion. The astrometric solutions were calculated using fits to local reference stars.}
\label{cpm}
\tiny
  \begin{tabular}{llcrcrccccl}
  \hline
Name&\multicolumn{1}{c}{Position}&$\mu$&P.A&$\sigma_{\mu}$&$\sigma_{PA}$&$Y$&$J$&$H$&$K$\\
&&''/yr&$^{\circ}$&''/yr&&&&\\
  \hline
ULAS2MASS1253+0740a&12 53 49.36 +07 40 04.5&0.162&269.6&0.026$^1$&5.6&14.559&13.807&13.323&12.870&\\ 
ULAS2MASS1253+0740b&12 53 49.69 +07 40 00.6&0.156&269.5&0.029$^1$&7.6&14.977&14.168&13.669&13.181&\\
   \hline
\normalsize
\end{tabular}
\end{minipage}
\end{table*}
\begin{table*}
 \begin{minipage}{160mm}
  \caption{Photometry and astrometry for objects in our sample which have a common proper motion companion found in another study. Astrometric solutions were calculated using fits to local reference stars. Some companion objects were brighter than the UKIDSS saturation limits. For these the photometry is drawn from 2MASS and they can be recognised by their lack of a $Y$ magnitude. Citation key - a: Tycho catalogue (Hog et al. 1998), b: Argelander (1903), c: Lepine \& Shara (2005), d: Hipparcos catalogue (Perryman 1997), e: Luyten (1979b), f: Giclas, Slaughter \& Thomas (1959), g: Hawley et al. (2002).}
\label{cpm1}
\tiny
  \begin{tabular}{lllllcrrrrl}
  \hline
Name&\multicolumn{1}{c}{Position}&$\mu$&P.A&$\sigma_{\mu}$&$\sigma_{PA}$&$Y$&$J$&$H$&$K$&note\\
&&''/yr&$^{\circ}$&''/yr&$^{\circ}$&&&&&\\
  \hline
ULAS2MASS0041+1341&00 41 54.44 +13 41 34.1&0.254&230.6&0.015&15.389&14.372&13.760&13.208&g\\ 
NLTT 2274&00 41 55.45 +13 41 16.4&0.268$^c$&228.5$^c$&0.008$^c$&1.7$^c$&&10.164&9.574&9.347&b\\
  \hline
ULAS2MASS0158-0025&01 58 05.95 -00 25 42.8&0.132&170.4&0.021$^1$&7.9&15.282&14.575&14.063&13.636&\\ 
BD-01 266&01 58 07.36 $-$00 25 10.8$^a$&0.141$^a$&172.9$^a$&0.002$^a$&1.1$^a$&&8.096&7.632&7.478&b\\
   \hline
ULAS2MASS0207+1355&02 07 35.73 +13 55 54.9&0.316&123.3&0.017$^1$&3.6&16.589&15.369&14.532&13.829&g\\ 
G 73-26&02 07 37.48 +13 54 49.5&0.321$^c$&125.4$^c$&0.008$^c$&1.4$^c$&&9.196&8.569&8.307&f\\
   \hline
ULAS2MASS0946+1116&09 46 12.09 +11 16 31.1&0.164&265.9&0.027$^1$&10.8&16.407&15.677&15.090&14.698&\\ 
NLTT 22538&09 46 12.64 +11 16 37.1&0.183$^c$&263.0$^c$&0.008$^c$&2.5$^c$&&10.233&9.698&9.554\\
   \hline
ULAS2MASS1159+0706&11 59 48.15 +07 06 59.1&0.201&301.1&0.023$^1$&6.2&16.027&15.304&14.801&14.399&\\ 
NLTT 29211&11 59 48.58 +07 07 08.6&0.204$^c$&302.4$^c$&0.008$^c$&2.3$^c$&17.606&17.526&17.575&17.491&e\\
   \hline
ULAS2MASS1259+0651&12 59 37.59 +06 51 18.9&0.444&240.3&0.032$^1$&5.0&15.045&14.289&13.851&13.422&\\
LSPM J1259+0651&12 59 39.34 +06 51 25.6$^c$&0.483$^c$&240.7$^c$&0.008$^c$&0.9$^c$&14.558&13.875&13.461&13.061&c\\
   \hline
ULAS2MASS1320+0957&13 20 41.49 +09 57 49.7&0.256&238.7&0.037$^1$&6.8&14.484&13.654&13.111&12.627&\\ 
NLTT 33793&13 20 50.13 +09 55 58.3$^d$&0.288$^d$&240.0$^d$&0.002$^d$&0.3$^d$&&7.889&7.323&7.218&e\\
   \hline
ULAS2MASS1327+0916&13 27 26.77 +09 16 05.6&0.146&238.3&0.012$^1$&5.6&15.300&14.539&13.994&13.578&\\
LSPM J1327+0916&13 27 28.51 +09 16 32.4$^c$&0.161$^c$&244.2$^c$&0.008$^c$&2.8$^c$&&10.68&10.07&9.83&c\\
   \hline
UGCS2MASS0336+2233&03 36 25.95 +22 33 17.2&0.243&132.7&0.007$^1$&1.7&15.808&14.991&14.471&13.977&\\
NLTT 11342&03 36 23.63 +22 33 27.3$^c$&0.244$^c$&134.5$^c$&0.008$^c$&1.9$^c$&&11.581&11.019&10.762&e\\
   \hline
UGCS2MASS0342+2248&03 42 10.17 +22 48 44.6&0.089&166.9&0.015$^1$&10.7&16.355&15.543&14.932&14.469&\\
HD 22992&03 42 15.92 +22 47 11.2$^d$&0.079$^d$&164.1&0.001$^d$&0.9$^d$&&6.512&6.343&6.313&d\\
   \hline
\normalsize
\end{tabular}
\end{minipage}
\end{table*}
We utilised the same technique that was used in Deacon \& Hambly (2007) to search for objects in our sample that may have a common proper motion companion within our sample. This means we searched for objects separated by less than three arcminutes with proper motions within two sigma of each other. This produced one potential pair, ULAS2MASS1253+0740a and ULAS2MASS1253+0740b. These are separated by only 6.3 arcseconds (only just outside the 2MASS proximity limit for good quality photometry and astrometry). The details of this pair can be seen in Table~\ref{cpm}. 

Additonally we checked all of our objects against the Simbad database to identify any from previous surveys which may be common proper motion companions. Using the same criteria as before we identified ten potential common proper motion pairs the details of which can be found in Table~\ref{cpm1}. Of these two have measured Hipparcos parallaxes, NLTT 33793 (a K8 star) with $\pi$=26.22$\pm$1.68 milli-arcseconds and HD 22992 (an F2 star) with $\pi$=18.79$\pm$1.05 milli-arcseconds. At these parallaxes, (with angular separations of 2.83 and 2.05 arcminutes respectively) both systems would have separations of roughly 6500AU. However in Section 3.4 we calculate distance estimates for these two objects. These lead us to conclude that HD 229920 and UGCS2MASS0342+2248 are not physically related but that NLTT 33793 and ULAS2MASS1320+0957 probably are. 
 \subsection{Spectroscopic Follow-up}
\begin{figure}
 \setlength{\unitlength}{1mm}
 \begin{picture}(75,120)
 \includegraphics{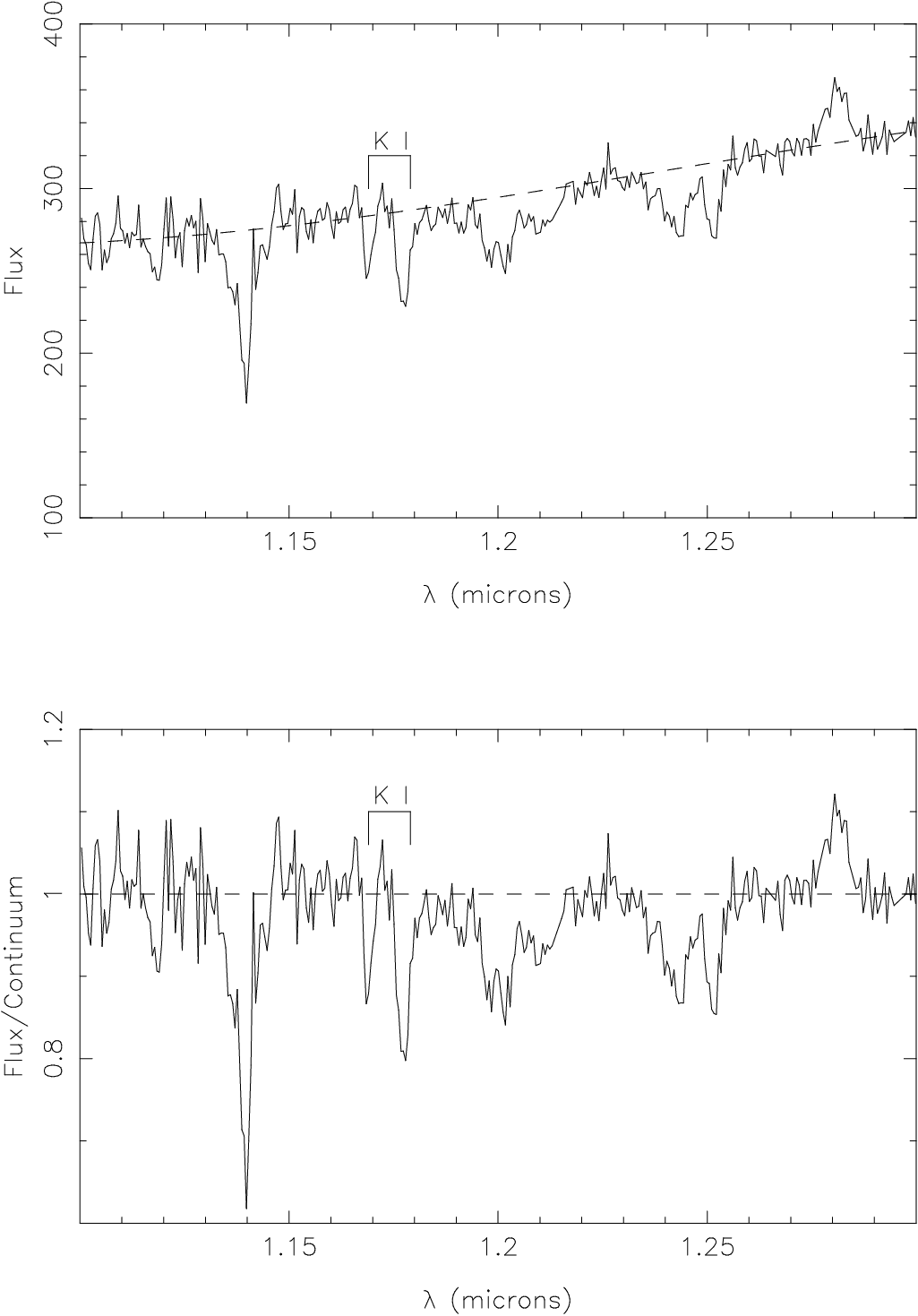}
 \end{picture}
 \caption[]{The infrared spectrum of ULAS2MASS1017+0719 also known as NLTT 23911. The top panel shows the flux calibrated spectrum and the lower panel the continuum divided spectrum. The spectral lines used for the classification are marked K I}
 \label{ULAS2MASS2}
 \end{figure}

\begin{figure}
 \setlength{\unitlength}{1mm}
 \begin{picture}(75,120)
 \includegraphics{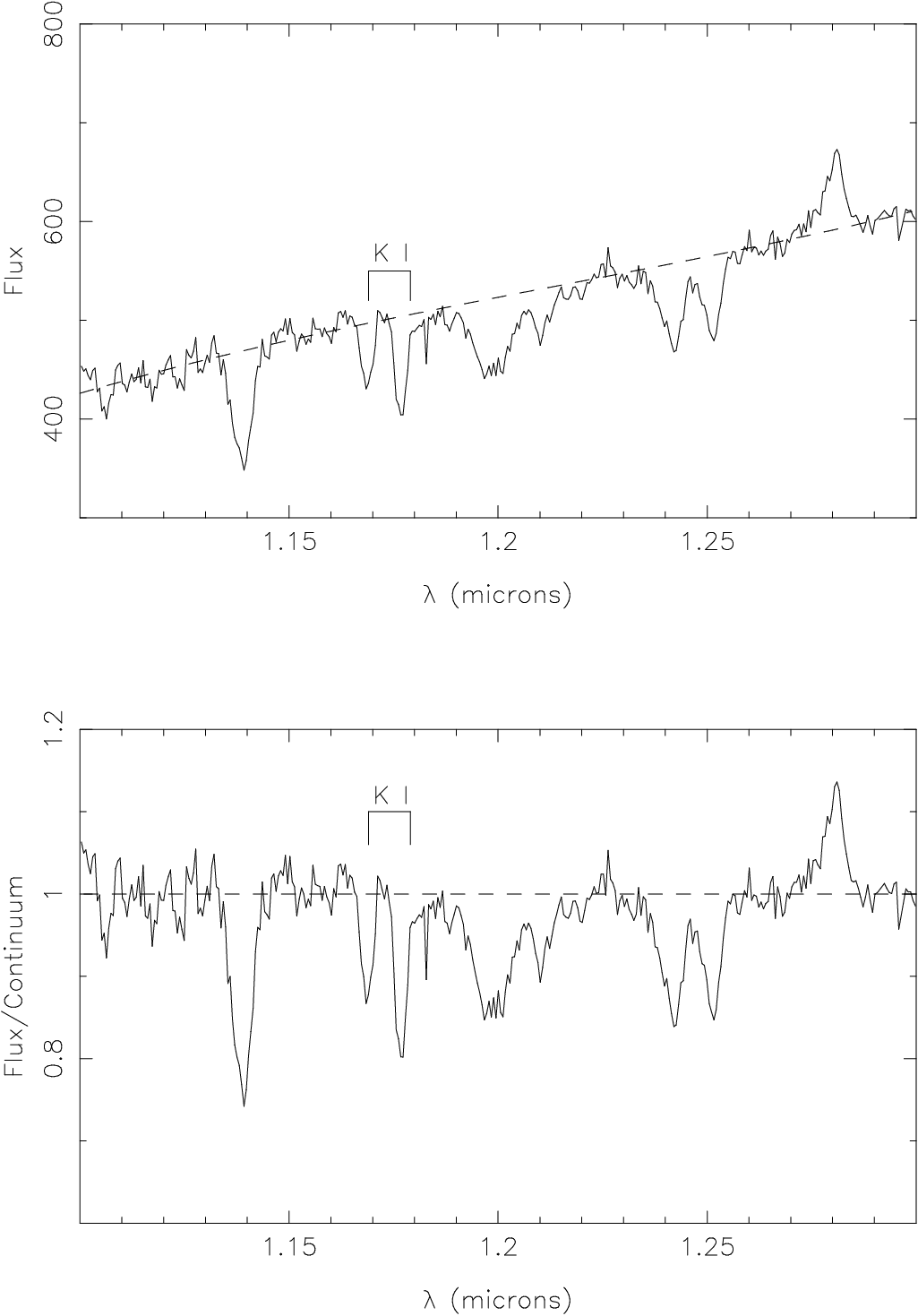}
 \end{picture}
 \caption[]{The infrared spectrum of ULAS2MASS1354+0846 also known as LSPM J1354+0846. The top panel shows the flux calibrated spectrum and the lower panel the continuum divided spectrum. The spectral lines used for the classification are marked K I}
 \label{ULAS2MASS4}
 \end{figure}

The sample presented here uses data from the UKIDSS Data Release 4. However before these data were available we used the same method on the smaller Data Release 2 to identify potentially interesting objects. As a result we applied for service observations on the UK Infrared Telescope (UKIRT) for a number of objects whose colours and magnitudes suggested they were within 30pc. Of these, two objects had spectra taken by UKIRT staff on the nights of the ninth and tenth of January 2008. These spectra were obtained using the IJ grating on the UIST spectrograph and consisted of four individual spectra jittered in an ABBA pattern. Standard star spectra and calibration frames were also taken. The data were reduced using the Figaro Starlink package. Figures~\ref{ULAS2MASS2} and~\ref{ULAS2MASS4} show the objects ULAS2MASS1017+0719 (previously known as NLTT 23911) and ULAS2MASS1354+0846 (previously known as LSPM J1354+0846) respectively. Table~\ref{spectra} shows the UKIDSS photometry and our calculated astrometry for the two objects.
\begin{table*}
 \begin{minipage}{160mm}
  \caption{The astrometry and photometry for the two objects for which spectra were taken. For both, the first line includes the astrometry and UKIDSS photometry and the second line the 2MASS photometry. No relative local astrometry was possible in the fields of these stars owing to a lack of suitable refernce stars, so the global astrometry was used for proper motion measurements. Citation key - a: Luyten NLTT, b: Lepine \& Shara (2005)}
\label{spectra}
\tiny
  \begin{tabular}{lccrcrccccl}
  \hline
Name&Position&$\mu$&P.A&$\sigma_{\mu}$&$\sigma_{PA}$&$Y$&$J$&$H$&$K$&note\\
&&''/yr&$^{\circ}$&''/yr&$^{\circ}$&&&&&\\
  \hline
ULAS2MASS1017+0711&10 17 23.35 +07 11 03.9&0.159&158.0&0.024$^1$&6.7&15.396&14.656&14.103&13.695&a\\ 
&&&&&&12.504&11.834&11.488&\\
ULAS2MASS1354+0846&13 54 08.67 +08 46 08.7&0.230&283.7&0.014$^2$&3.2&12.976&12.146&11.587&11.144&b\\ 
&&&&&&12.193&11.605&11.156&\\
  \hline
\normalsize
\end{tabular}
\end{minipage}
\end{table*}
We used the equivalent widths of K I lines at 1.169 and 1.179 microns and the values for those lines quoted in Cushing et al. (2005) for each spectral type to determine spectral classification. With an EW (in angstr\"oms) of 6.9 for the 1.179$\mu$m line and 2.4 for the 1.169$\mu$m line we deduce that ULAS2MASS1017+0719 is of spectral type M6-8, most probably M7. With equivalent widths of 7.1 and 3.5 respectively we estimate that ULAS2MASS1354+0846 is of the same spectral type. If we compare the 2MASS photometry of our objects with that of the M7 dwarf SO0253+1652 (which Henry et al., 2006 found to lie at 3.85$\pm^{0.03}_{0.05}$pc) we can produce a rough distance estimate for both objects. SO0253+1652 has 2MASS magnitudes $J$=8.394, $H$=7.883 and $K_s$=7.585. Based on this photometry we estimate that ULAS2MASS1017+0719 lies at roughly 23.7pc and ULAS2MASS1354+0846 at 21.0pc. Clearly these are rough estimates and are inferior to trigonometric parallax measurements.  
\subsection{Photometric Distance Estimates}

\begin{figure}
 \setlength{\unitlength}{1mm}
 \begin{picture}(75,230)
 \includegraphics{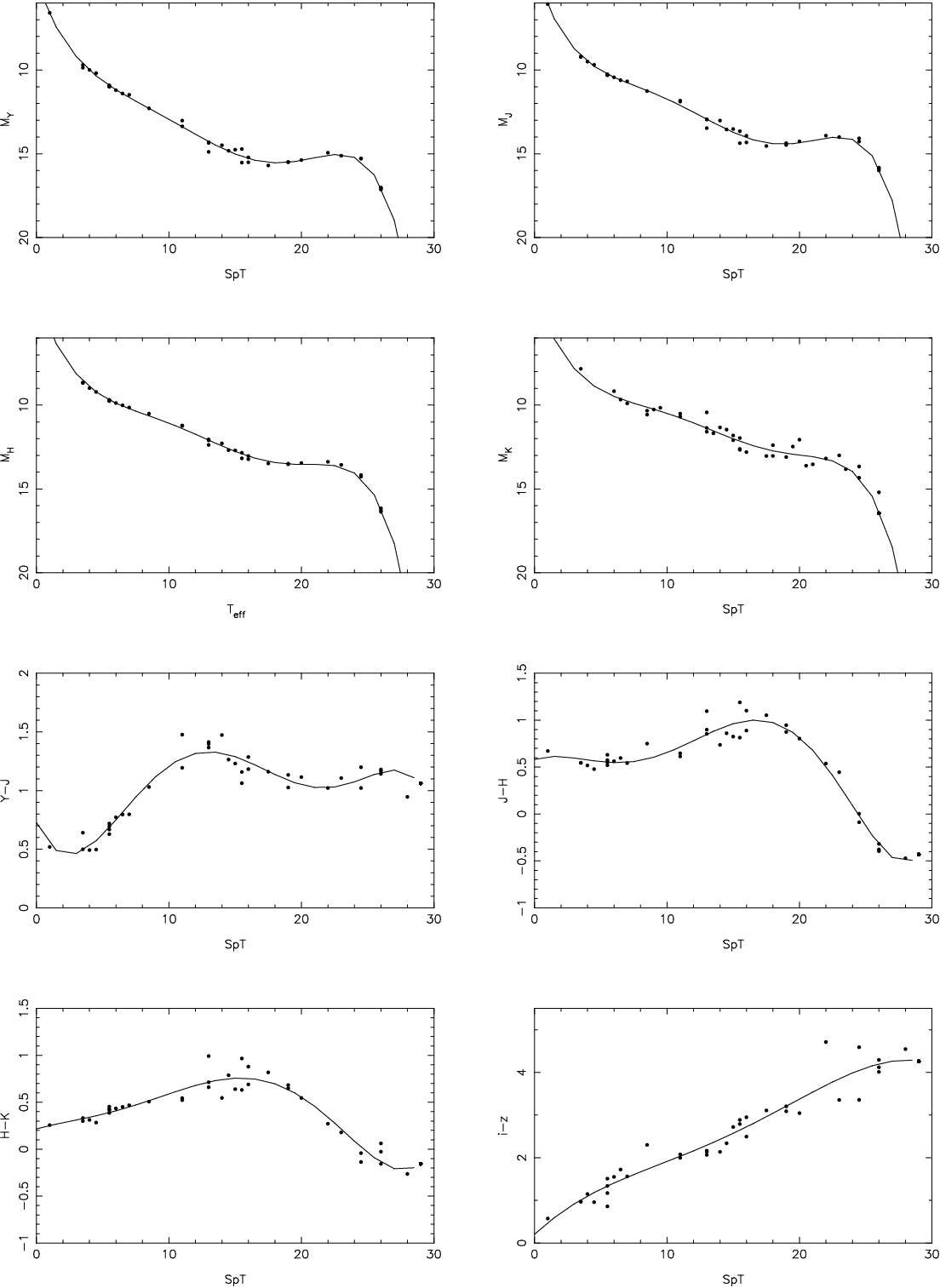}
 \end{picture}
 \caption[]{Plots of the data and fits for the colour magnitude relations. The data here includes the colours from Hewett et al. (2006) and absolute magnitudes from Golimowski et al. (2004). M0 is spectral type 0, L0 spectral type 10 and T0 spectral type 20.}
 \label{colours}
 \end{figure}

While a list of potential nearby cool dwarfs is useful, photometric distance estimates can help to identify the nearest objects in our sample. To do this we took the data from Hewett et al. (2006) and Leggett (private communication) for the WFCAM colours of cool dwarfs and combined these data with those from Golimowski et al. (2004) (which is itself based on a series of trigonometric parallax measurements from the literature) to produce a series of relations. The first set of these relate colours to spectral type. These are plotted in the lower panels of Figure~\ref{colours}. Separate polynomials for the M and L and T dwarf regimes are fitted, these are shown in Table~\ref{relations}. The $Y-J$, $J-H$ and $H-K$ colours were used to calculate a separate spectral type for each for the target object. As many of these relations have no unique solution for some colours, more than one spectral type for each colour was normally calculated. Next the spectral types were grouped together so that three spectral types from the three colours were in the close agreement. In some cases there was more than one valid grouping of estimated spectral types. Each group was then averaged to calculate the estimated spectral type. If there is more than one calculated spectral type, other data must be used to select the correct estimation. Here we use $i$ and $z$ photometry from the Sloan Digital Sky Survey. The fourth colour relation we use plots $i-z$ vs. spectral type. We use this relation along with Sloan photometry and choose the calculated spectral type where the photometry and the colour relation are in best agreement. Once this spectral type was calculated the observed magnitudes in each passband could be compared with the relation between spectral type and the absolute magnitude in that particular passband (see Table~\ref{relations} and the top 4 panels of Figure~\ref{colours}). A distance from each passband was then calculated and these were averaged to produce a final distance estimate.

Clearly the relations must be tested to ensure they are reliable. To do this we first took the measured spectral types for our potential L and T dwarfs (see Table~\ref{Ldwarfs}). We then compared these with each object's spectral calculated by our method. We found that in general we had a standard deviation of roughly 1.5 subtypes. We also used objects common between our sample and Reid et al. (2008) to estimate the accuracy of our distance estimates. These had calculated photometric distances from 2MASS data and from Cruz et al. (2003). We compared these distances with our own and found that (once the errors on the Reid et al. distance calculations had been taken into account) our calculated distances were accurate to roughly 20\%. We recommend that these relations are only used for initial sample selection. We do not believe these are accurate enough to quote values for distances. Table~\ref{close} shows a list of all the objects in our sample which we believe to be within 30pc based on our photometric distance relations.  Our sample includes both the objects which we spectroscopically classified. In the previous section we calculated the distances to these objects using an independent method, both were estimated to be M7 dwarfs at roughly 20pc. Here, both have distance estimates of roughly 20pc with one estimated to be an M7 dwarf and one an M6.5 dwarf. In Section 3.2 we identified two objects which shared a common proper motion with nearby stars with Hipparcos parallaxes. Using the photometric relations we find that UGCS2MASS0342+2248 lies at roughly 80.2 parsecs and ULAS2MASS1320+0957 at approximately 38.2 parsecs. As HD229920 - the common proper motion companion of UGCS2MASS0342+2248 - lies at 53pc, we must conclude these objects are unlikely to be related. However NLTT 33793 lies at 38.2pc, a good match to our distance estimate for ULAS2MASS1320+0957. 
\begin{table*}
 \begin{minipage}{160mm}
  \caption{The colour/magnitude to effective temperature relations used in this work. They are of the form $X=\sum_i a_i SpT^i$ where M0 is spectral type 0, L0 spectral type 10 and T0 spectral type 20. }
\label{relations}
\footnotesize
\begin{center}
  \begin{tabular}{lccccccc}
  \hline
Relation&$a_0$&$a_1$&$a_2$&$a_3$&$a_4$&$a_5$&$a_6$\\
  \hline
$M_Y$&4.800e+00&2.147e+00&-2.844-01&2.035e-02&-4.502e-04&-1.167e-05&4.384e-07\\ 
$M_J$&4.074e+00&2.394e+00&-3.506e-01&2.488e-02&-5.166e-04&-1.447e-05&5.090e-07\\ 
$M_H$&3.492e+00&2.345e+00&-3.291e-01&2.233e-02&-4.442e-04&-1.304e-05&4.513e-07\\ 
$M_K$&3.276e+00&2.306e+00&-3.243e-01&2.154e-02&-4.130e-04&-1.261e-05&4.277e-07\\ 
$Y-J$&7.255e-01&-2.470e-01&6.613e-02&-4.531e-03&6.634e-05&2.802e-06&-7.063e-08\\
$J-H$&5.81e-01&4.888e-02&-2.151e-02&2.543e-03&-7.233e-05&-1.440e-06&5.768e-08\\
$H-K$&2.167e-01&3.950e-02&-4.808e-03&7.891e-04&-3.127e-05&-4.204e-07&2.358e-08\\ 
$i-z$&2.037e-01&2.891e-01&-1.951e-02&7.997e-04&2.458e-06&-6.175e-07&5.917e-09\\ 
  \hline
\normalsize
\end{tabular}
\end{center}
\end{minipage}
\end{table*}
\begin{table*}
 \begin{minipage}{160mm}
  \caption{Objects with photometric distance estimates within 30pc according to our photometric distance relations. The first spectral type and distance are calculated using our distance estimates, the second spectral type and distance estimate come from another source. Citation key - a: Burgasser et al. (2004), b: Burgasser et al. (2006), c: Wilson et al. (2003), d: Reid et al. (2008), e: Tinney (1993), f: Tinney (1995), g: Lepine \& Shara (2005), h: this work, i: Hawley et al. (2002), j: Luyten (1979b), k: Kirkpatrick et al. (1997), l: Cruz et al. (2003), m: Knapp et al. (2004), n: Phan-Bao \& Bessell (2006), o: Kirkpatrick et al. (2000).}
\label{close}
\footnotesize
  \begin{tabular}{lcccccccccccl}
  \hline
Name&$\mu$&$Y$&$J$&$H$&$K$&SpT&d&SpT&d&note\\
&''/yr&&&&&&(pc)&&(pc)&\\
  \hline
ULAS2MASS1231+0847&1.590&16.350&15.166&15.451&15.544&T5.5&9.1&T5.5&&a,b\\
ULAS2MASS1448+1031&0.278&15.813&14.425&13.520&12.695&L4&15.9&L4&19.6$\pm$4.0&c,d\\
ULAS2MASS1510-0241&0.399&13.486&12.534&11.974&11.313&M9&16.7&M9&16.34$\pm$1.25&e,d,f\\
ULAS2MASS0002+0115&0.455&12.894&12.093&11.545&11.074&M6.5&19.3&M6.5&21.0$\pm$2.52&j,n\\
ULAS2MASS0001+1535&0.218&16.877&15.462&14.483&13.623&L5&20.8&L4&&m\\
ULAS2MASS1354+0846&0.230&12.979&12.151&11.600&11.153&M7&21.6&M7&&g,h\\
ULAS2MASS1134+0022&0.511&13.689&12.790&12.181&11.701&M8&21.7&M9&18.9$\pm$1.2&i,d\\
ULAS2MASS1017+0719&0.188&13.213&12.407&11.872&11.424&M6.5&22.6&M7&&j,h\\
ULAS2MASS0205+1251&0.362&17.018&15.561&14.513&13.637&L5&22.8&L5&26.6$\pm$3.0&o,l\\
UGCS2MASS0339+2457&0.178&13.608&12.772&12.223&11.732&M7&23.5&M8&&k\\
ULAS2MASS1407+1241&0.337&16.712&15.319&14.381&13.617&L4.5&23.6&L1&47.4$\pm$12.5&d\\
UGCS2MASS0409+2104&0.184&16.622&15.376&14.502&13.781&L4&24.8&L3&&o\\
UGCS2MASS0354+2316&0.206&13.903&13.054&12.469&11.982&M7.5&25.5&&&g\\
ULAS2MASS1457+1102&0.117&13.169&12.466&12.013&11.616&M6&26.5&&&\\
ULAS2MASS1546+0317&0.113&13.965&13.155&12.582&12.114&M7&27.7&\\
UGCS2MASS0345+2540&0.109&15.009&13.922&13.246&12.663&M9.5&28.4&L0&26.95$\pm$0.36&k,d\\
  \hline
\normalsize
\end{tabular}
\end{minipage}
\end{table*}
\section{Discussion}
\begin{figure}
 \setlength{\unitlength}{1mm}
 \begin{picture}(75,130)
 \includegraphics{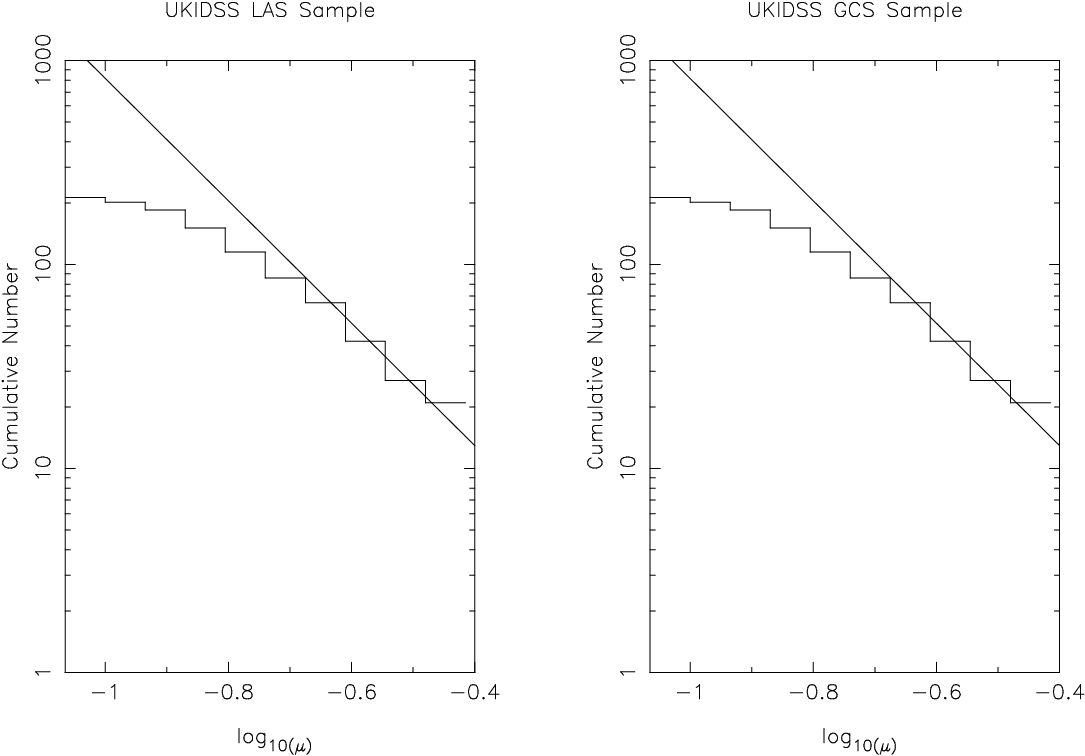}
 \end{picture}
 \caption[]{Cumulative histograms for both samples. With a survey with no incompleteness the number should scale as $\mu^{-3}$ (indicated by the solid line). It is clear that below 0.2''/yr our surveys deviates from this trend and is hence incomplete.}
 \label{muhist}
 \end{figure}

To provide a simple estimate of completeness we plotted a cumulative histogram of objects' proper motions. Assuming the stellar velocity and positional distributions do not change significantly, the cumulative number of objects should scale as $\mu^{-3}$ (see Figure~\ref{muhist}). A line showing this scaling is also plotted. It is clear that the completeness begins to drop off below about $\mu=0.2$''/yr. This will be due to distant, fainter objects being excluded from our sample due to magnitude limits. At the lowest proper motions higher astrometic errors may also cause some objects to be excluded. To check our completeness for cool objects we used the Dwarf Archives website to identify any potential L and T dwarfs in our survey area with brightnesses within our detection range. Eleven L dwarfs of sufficient brightness were identified in our survey area. Of these one (GD 165B) did not appear in our sample. The reason for this is that the 2MASS proximity flag for this object was below six arcseconds due to the closeness of its binary companion. Of the known T dwarfs only 2MASS12314753+0847331 lies in the survey area, this object is detected. Hence we detect all the known, single L and T dwarfs brighter than the 2MASS limit in our survey area

In order to examine the accuracy of our proper motions we cross--referenced our catalogue with that of Lepine \& Shara (2005). In the LAS sample there were fifteen common objects between the two surveys. Our proper motions of these agreed well with those of Lepine and Shara. However there were large offsets found in two of the four objects in both Lepine and Shara and our GCS sample. The reason for this is that we are measuring relative proper motions while Lepine \& Shara measure absolute proper motions. Put simply both our relative astrometry and the UKIDSS astrometry use stars to define our reference frame, Lepine and Shara use quasars to set their astrometric reference frame. This leaves us vulnerable to the proper motions of our reference stars introducing an error to our astrometric system. As the Galactic Cluster Survey is taken in the area of known open clusters, there will be a distinct bulk motion of a large number of potential reference stars. Hence there will be an offset in our proper motions measurements.

The objects found in the UKIDSS GCS are probably unrelated to the target clusters. As we have chosen a minimum proper motion of 80 milliarcseconds per year all our objects have proper motions at least 2-3 $\sigma$ higher than the proper motions of nine out of the ten target clusters and star forming regions. The one exception is the Hyades, however the core area of the Hyades does not yet have the UKIDSS GCS multiband observations required for this survey. Additionally several objects are listed as being cluster members in Table~\ref{GCSobjects} (this cluster membership flagging comes from their descriptions in SIMBAD). These objects are listed as either members of the Pleiades or $\alpha$ Per. If our proper motion measurements for these objects are correct it would indicate that they are not members of these clusters. 

 \section{Conclusions}
Our attempt to utilise the currently available UKIDSS data for a proper motion survey has successfully detected all the expected, previously known single L and T dwarfs in our LAS survey area. Additionally we have found seven objects in the LAS and one in the GCS that are good candidate L dwarfs. Finally, initial spectral follow--up of our sample has classified two previously known objects. Rough distance estimates put these as being within 25pc. In further iterations of this work with future UKIDSS data releases we will attempt to tie our proper motions to absolute proper motions, avoiding the offsets in the Galactic Clusters Survey sample. This work could easily be extended to cover the upcoming surveys by the VISTA telescope.

\section*{Acknowledgments}
N.R.D. is supported by NWO--VIDI grant 639.042.201 to P. J. Groot. This research has made use of the SIMBAD database, operated at CDS, Strasbourg, France and of DwarfArchives.org, maintained by Chris Gelino, Davy Kirkpatrick, and Adam Burgasser. The UKIDSS project is defined in Lawrence et al (2007). UKIDSS uses the UKIRT Wide Field Camera (WFCAM; Casali et al 2007) and a photometric system described in Hewett et al (2006). The pipeline processing and science archive are described in Irwin et al (2008) and Hambly et al (2008). We have used data from the 2nd, 3rd and 4th data releases, the first of which is described in detail in Warren et al. (2007b). This publication makes use of data products from the Two Micron All Sky Survey, which is a joint project of the University of Massachusetts and the Infrared Processing and Analysis Center/California Institute of Technology, funded by the National Aeronautics and Space Administration and the National Science Foundation. The authors would like to thank the UKIRT staff for carrying out service observations used in this publication. The Slalib positional astronomy library by Wallace (1998) was used in this work.

\appendix
\section{SQL Query used} 
\small
\begin{verbatim}
SELECT las.ra as alpha, las.dec as delta, las.yAperMag3, las.j_1AperMag3, las.hAperMag3, las.kAperMag3, mj.mjdObs
FROM  
      lasSource AS las, Multiframe AS mj, lasMergeLog AS l

WHERE j_1mfID=mj.multiframeID
      AND las.frameSetID=l.frameSetID
      AND yClassStat   BETWEEN -3.0 AND 3.0
      AND j_1ClassStat BETWEEN -3.0 AND 3.0
      AND hClassStat   BETWEEN -3.0 AND 3.0
      AND kClassStat   BETWEEN -3.0 AND 3.0
      AND j_1Xi BETWEEN -1.0 AND +1.0
      AND hXi BETWEEN -1.0 AND +1.0
      AND kXi BETWEEN -1.0 AND +1.0
      AND j_1Eta BETWEEN -1.0 AND +1.0
      AND hEta BETWEEN -1.0 AND +1.0
      AND kEta BETWEEN -1.0 AND +1.0
      AND las.sourceID NOT IN (SELECT masterObjID FROM lasSourceXtwomass_psc WHERE distancemins < 0.01)
      AND (las.priOrSec=0 OR las.priOrSec=las.frameSetID)
      AND (las.yAperMag3-las.j_1AperMag3)>0.3
      AND ((las.yAperMag3-las.j_1AperMag3)>0.7 OR (las.j_1AperMag3-las.hAperMag3)<0.0) 
      AND las.j_1AperMag3<16.0 
      AND las.j_1AperMag3>10.5
      AND las.yAperMag3>11.3
      AND las.hAperMag3>10.2
      AND las.kAperMag3>9.7
      AND mj.mjdObs>0.0
      AND las.j_1Ell<0.3
      AND las.yEll<0.3
      AND las.hEll<0.3
      AND las.kEll<0.3
\end{verbatim}
\normalsize
\section{Data Tables} 
\begin{table*}
   \caption{\label{LASobjects} Objects identified in our LAS sample. All photometry is in the UKIDSS system (Hewett et al. 2006). $^1$ denotes objects whose astrometric solutions was calculated using local reference stars while $^2$ denotes those calculated using only global error estimates. Citation key - a:Schneider et al. (2002) b: Luyten (1979b) c: Tinney (1993) d: Fan et al. (2000) e: Bouy et al. (2003) f: Lepine \& Shara (2005) g: Cruz et al. (2003) h: Hawley et al. (2002) i: Delfosse et al. (1999) j: Burgasser et al. (2004) k: Luyten (1979a) l Cruz et al. (2007) m: Wilson et al. (2003) n: Reid et al. (2008), p: Knapp et al. (2004), q: Gizis et al. (2000), r:Kirkpatrick et al. (2000).}
\tiny
\begin{tabular}{lccrcrccccl}
\hline 
Name&Position&$\mu$&P.A&$\sigma_{\mu}$&$\sigma_{PA}$&$Y$&$J$&$H$&$K$&note\\
&&''/yr&$^{\circ}$&''/yr&$^{\circ}$&&&&&\\
  \hline
ULAS2MASS0001+1545&00 01 02.13 +15 45 51.9&0.375&30.8&0.015$^1$&2.1&14.781&14.027&13.653&13.209&\\ 
ULAS2MASS0001+1535&00 01 12.24 +15 35 34.3&0.218&144.0&0.029$^1$&8.4&16.877&15.462&14.483&13.623&p\\ 
ULAS2MASS0001+1544&00 01 47.80 +15 44 11.0&0.108&127.4&0.017$^1$&11.2&15.808&15.089&14.557&14.160&\\ 
ULAS2MASS0002+0115&00 02 06.45 +01 15 36.7&0.455&79.9&0.013$^2$&1.4&12.894&12.093&11.545&11.074&b\\ 
ULAS2MASS0029+1509&00 29 16.74 +15 09 17.8&0.119&100.4&0.021$^1$&10.1&16.343&15.589&15.176&14.769&\\ 
ULAS2MASS0033-0015&00 33 05.57 $-$00 15 29.7&0.148&184.9&0.011$^1$&8.4&14.224&13.517&13.028&12.654&\\ 
ULAS2MASS0041+1341&00 41 54.44 +13 41 34.1&0.254&230.6&0.027$^1$&5.3&15.389&14.372&13.760&13.208&h\\ 
ULAS2MASS0054-0031&00 54 06.66 $-$00 31 03.2&0.248&129.8&0.029$^1$&6.9&16.857&15.629&14.917&14.302&a\\ 
ULAS2MASS0054-0009&00 54 09.96 $-$00 09 15.9&0.120&79.3&0.015$^1$&5.8&15.074&14.372&13.836&13.461&\\ 
ULAS2MASS0104+1457&01 04 37.80 +14 57 24.3&0.158&88.0&0.009$^2$&2.8&14.467&13.556&13.017&12.545&q\\ 
ULAS2MASS0106+1518&01 06 37.14 +15 18 53.5&0.407&238.6&0.008$^2$&1.1&15.095&14.277&13.867&13.363&\\ 
ULAS2MASS0130+1319&01 30 25.84 +13 19 47.1&0.154&209.7&0.016$^1$&4.9&15.023&14.301&13.744&13.344&\\ 
ULAS2MASS0135+1307&01 35 00.62 +13 07 56.6&0.218&117.5&0.022$^1$&7.6&15.771&15.048&14.545&14.124&\\ 
ULAS2MASS0151+1412&01 51 19.74 +14 12 03.8&0.071&77.7&0.010$^1$&10.1&14.954&14.242&13.662&13.232&\\ 
ULAS2MASS0152+0722&01 52 38.80 +07 22 07.6&0.139&109.9&0.020$^1$&5.3&15.734&14.923&14.302&13.896&\\ 
ULAS2MASS0158-0025&01 58 05.95 $-$00 25 42.8&0.132&170.4&0.021$^1$&7.9&15.282&14.575&14.063&13.636&\\ 
ULAS2MASS0200+0038&02 00 23.65 +00 38 46.5&0.160&120.0&0.024$^1$&7.2&15.139&14.389&13.847&13.402&c\\ 
ULAS2MASS0205+1251&02 05 03.66 +12 51 42.0&0.362&93.8&0.025$^1$&4.3&17.018&15.561&14.513&13.637&r\\ 
ULAS2MASS0205-0012&02 05 17.71 $-$00 12 34.4&0.129&106.3&0.025$^1$&17.7&16.557&15.720&15.134&14.672&\\ 
ULAS2MASS0207+1355&02 07 35.73 +13 55 54.9&0.316&123.3&0.017$^1$&3.6&16.589&15.369&14.532&13.829&p\\ 
ULAS2MASS0212+1404&02 12 18.71 +14 04 07.4&0.108&232.9&0.015$^1$&7.7&16.023&15.219&14.620&14.162&\\ 
ULAS2MASS0212+0559&02 12 29.24 +05 59 05.5&0.305&102.6&0.012$^1$&2.4&15.056&14.346&13.763&13.343&\\ 
ULAS2MASS0215+0015&02 15 46.88 +00 15 28.0&0.271&76.2&0.022$^1$&4.9&16.313&15.499&14.888&14.443&\\ 
ULAS2MASS0219+0506&02 19 22.05 +05 06 30.8&0.184&84.1&0.013$^1$&5.5&16.022&14.855&14.089&13.453&\\ 
ULAS2MASS0233+0031&02 33 21.47 +00 31 33.3&0.130&138.6&0.022$^1$&9.3&16.329&15.547&15.001&14.531&\\ 
ULAS2MASS0249-0101&02 49 49.23 $-$01 01 58.0&0.135&150.3&0.027$^1$&10.2&16.157&15.389&14.819&14.424&\\ 
ULAS2MASS0250+0532&02 50 09.71 +05 32 23.9&0.129&141.4&0.019$^1$&8.9&15.312&14.479&13.900&13.472&\\ 
ULAS2MASS0251+0047&02 51 13.35 +00 47 36.8&0.242&72.6&0.017$^2$&3.7&14.549&13.713&13.115&12.632&c\\ 
ULAS2MASS0304+0045&03 04 02.18 +00 45 51.4&0.257&83.4&0.018$^2$&2.8&12.461&11.754&11.275&10.906&c\\ 
ULAS2MASS0324-0050&03 24 53.92 $-$00 50 08.9&0.106&109.0&0.017$^1$&9.3&15.508&14.755&14.237&13.793&\\ 
ULAS2MASS0330-0025&03 30 35.32 $-$00 25 37.4&0.509&131.1&0.015$^1$&1.7&16.508&15.215&14.487&13.771&d\\ 
ULAS2MASS0344+0111&03 44 08.88 +01 11 23.9&0.209&211.1&0.030$^1$&6.1&15.698&14.554&13.952&13.446&e\\ 
ULAS2MASS0346-0048&03 46 11.02 $-$00 48 29.2&0.142&145.6&0.013$^2$&5.5&16.348&15.606&15.148&14.724&\\ 
ULAS2MASS0818+0613&08 18 20.71 +06 13 26.9&0.118&319.0&0.010$^1$&4.4&14.530&13.762&13.210&12.775&\\ 
ULAS2MASS0819+0636&08 19 29.62 +06 36 54.6&0.172&217.8&0.016$^1$&4.9&14.644&13.936&13.429&13.023&f\\ 
ULAS2MASS0819+0453&08 19 47.89 +04 53 10.8&0.166&301.0&0.013$^1$&5.4&14.069&13.366&12.805&12.451&b\\ 
ULAS2MASS0821+0713&08 21 11.94 +07 13 00.1&0.153&170.0&0.026$^1$&9.0&16.442&15.635&15.129&14.686&\\ 
ULAS2MASS0821+0651&08 21 12.61 +06 51 06.8&0.092&229.8&0.013$^1$&8.2&15.844&14.990&14.332&13.832&\\ 
ULAS2MASS0829-0012&08 29 48.97 $-$00 12 22.8&0.266&309.2&0.010$^1$&2.2&14.268&13.405&12.850&12.400&g\\ 
ULAS2MASS0835+0000&08 35 09.59 +00 00 53.8&0.189&115.8&0.010$^1$&2.8&14.773&13.966&13.433&12.986&\\ 
ULAS2MASS0835+0548&08 35 58.24 +05 48 30.7&0.110&257.6&0.009$^1$&7.7&15.705&14.499&13.762&13.127&n\\ 
ULAS2MASS0836+0221&08 36 13.43 +02 21 06.5&0.108&289.9&0.012$^1$&7.8&15.718&14.706&14.164&13.656&\\ 
ULAS2MASS0836+0937&08 36 27.22 +09 37 17.7&0.129&157.1&0.018$^1$&10.6&15.916&15.214&14.664&14.279&\\ 
ULAS2MASS0836-0004&08 36 52.61 $-$00 04 32.4&0.150&141.8&0.009$^1$&3.6&15.993&15.209&14.635&14.215&\\ 
ULAS2MASS0837-0104&08 37 38.74 $-$01 04 39.6&0.211&279.6&0.023$^1$&6.6&16.298&15.397&14.855&14.377&\\ 
ULAS2MASS0842+1026&08 42 52.25 +10 26 39.0&0.189&221.3&0.027$^1$&8.0&15.371&14.603&14.037&13.601&\\ 
ULAS2MASS0843+1024&08 43 33.31 +10 24 43.0&0.593&165.3&0.011$^1$&1.5&15.989&14.777&14.105&13.551&\\ 
ULAS2MASS0844+0434&08 44 03.38 +04 34 37.3&0.426&310.3&0.028$^1$&4.1&14.309&13.394&12.845&12.359&\\ 
ULAS2MASS0846+0614&08 46 08.50 +06 14 47.9&0.090&257.0&0.011$^1$&10.4&15.450&14.725&14.223&13.787&\\ 
ULAS2MASS0858-0217&08 58 12.46 $-$02 17 32.9&0.154&271.8&0.020$^1$&5.6&16.172&15.219&14.597&14.069&\\ 
ULAS2MASS0900+0556&09 00 32.56 +05 56 24.4&0.143&280.1&0.011$^1$&10.9&15.793&14.951&14.421&13.920&\\ 
ULAS2MASS0908+0836&09 08 04.60 +08 36 45.2&0.121&155.5&0.017$^1$&6.3&14.988&14.059&13.451&12.947&\\ 
ULAS2MASS0908+1102&09 08 31.18 +11 02 25.3&0.210&244.9&0.011$^1$&3.4&15.105&14.364&13.815&13.389&\\ 
ULAS2MASS0908+1113&09 08 51.74 +11 13 24.1&0.102&268.3&0.019$^1$&8.1&16.325&15.554&15.050&14.664&\\ 
ULAS2MASS0911+0703&09 11 10.45 +07 03 38.9&0.133&194.2&0.015$^1$&9.7&15.063&14.351&13.806&13.405&\\ 
ULAS2MASS0911+0023&09 11 34.07 +00 23 12.5&0.211&269.9&0.033$^1$&6.8&13.827&13.066&12.531&12.131&b\\ 
ULAS2MASS0915+0606&09 15 13.58 +06 06 30.2&0.152&282.5&0.028$^1$&12.8&16.356&15.597&15.135&14.743&\\ 
ULAS2MASS0919+0016&09 19 42.72 +00 16 03.3&0.227&154.0&0.014$^1$&3.0&14.779&13.906&13.258&12.783&\\ 
ULAS2MASS0919-0004&09 19 49.65 $-$00 04 14.5&0.144&150.2&0.017$^1$&6.7&15.343&14.636&14.097&13.689&\\ 
ULAS2MASS0920+0640&09 20 50.85 +06 40 33.9&0.141&246.2&0.020$^1$&6.3&16.158&15.310&14.756&14.319&\\ 
ULAS2MASS0926-0108&09 26 05.60 $-$01 08 23.4&0.187&219.0&0.011$^1$&3.2&14.511&13.778&13.232&12.843&\\ 
ULAS2MASS0926-0118&09 26 09.36 $-$01 18 51.2&0.128&198.4&0.022$^1$&9.3&15.986&15.051&14.486&13.992&\\ 
ULAS2MASS0927+0045&09 27 09.24 +00 45 05.0&0.101&260.6&0.016$^1$&11.3&15.213&14.456&13.903&13.493&\\ 
ULAS2MASS0929+1018&09 29 01.77 +10 18 25.5&0.115&244.3&0.017$^1$&7.4&14.987&14.259&13.728&13.325&\\ 
ULAS2MASS0934-0024&09 34 22.97 $-$00 24 07.7&0.240&262.7&0.012$^1$&4.4&14.508&13.794&13.271&12.864&\\ 
ULAS2MASS0936-0053&09 36 30.49 $-$00 53 06.3&0.145&306.6&0.026$^1$&9.2&16.165&15.162&14.550&14.015&a\\ 
ULAS2MASS0936+0614&09 36 33.89 +06 14 57.5&0.126&147.7&0.019$^1$&9.4&15.499&14.521&13.905&13.345&\\ 
ULAS2MASS0937+0704&09 37 31.88 +07 04 15.0&0.150&156.9&0.024$^1$&11.9&16.368&15.413&14.787&14.274&\\ 
ULAS2MASS0941+1009&09 41 34.90 +10 09 41.1&0.280&190.9&0.016$^1$&4.0&15.547&14.514&13.900&13.364&\\ 
ULAS2MASS0946+1116&09 46 12.09 +11 16 31.1&0.164&265.9&0.027$^1$&10.8&16.407&15.677&15.090&14.698&\\ 
ULAS2MASS0947+0617&09 47 50.60 +06 17 13.0&0.274&236.7&0.028$^1$&5.0&15.371&14.397&13.708&13.167&\\ 
ULAS2MASS0952-0303&09 52 40.08 $-$03 03 37.4&0.220&267.2&0.017$^1$&4.0&15.045&14.254&13.699&13.278&\\ 
ULAS2MASS0954+0821&09 54 06.49 +08 21 33.9&0.217&251.3&0.028$^1$&3.0&14.746&13.906&13.391&12.941&\\ 
ULAS2MASS0956+0933&09 56 00.04 +09 33 27.8&0.120&215.9&0.021$^1$&8.2&15.624&14.708&14.041&13.544&\\ 
ULAS2MASS0957-0151&09 57 56.06 $-$01 51 21.2&0.230&301.6&0.040$^1$&8.9&16.554&15.728&15.210&14.752&\\ 
ULAS2MASS1001-0228&10 01 52.85 $-$02 28 08.4&0.161&174.5&0.022$^1$&9.9&15.866&15.160&14.666&14.258&\\ 
ULAS2MASS1003-0035&10 03 05.04 $-$00 35 09.5&0.152&233.6&0.022$^1$&8.0&15.786&14.816&14.163&13.637&\\ 
ULAS2MASS1009+0905&10 09 40.74 +09 05 48.6&0.472&259.1&0.023$^1$&1.5&15.258&14.472&14.013&13.639&\\ 
ULAS2MASS1012+1105&10 12 15.76 +11 05 47.4&0.097&287.5&0.017$^1$&14.1&16.218&15.481&14.913&14.520&\\
 \hline
\end{tabular}
\end{table*}
\addtocounter{table}{-1}
\begin{table*}
   \caption{\label{LASobjects} Objects identified in our LAS sample. All photometry is in the UKIDSS system (Hewett et al. 2006). $^1$ denotes objects whose astrometric solutions was calculated using local reference stars while $^2$ denotes those calculated using only global error estimates. Citation key - a:Schneider et al. (2002) b: Luyten (1979b) c: Tinney (1993) d: Fan et al. (2000) e: Bouy et al. (2003) f: Lepine \& Shara (2005) g: Cruz et al. (2003) h: Hawley et al. (2002) i: Delfosse et al. (1999) j: Burgasser et al. (2004) k: Luyten (1979a) l Cruz et al. (2007) m: Wilson et al. (2003) n: Reid et al. (2008), p: Knapp et al. (2004), q: Gizis et al. (2000), r:Kirkpatrick et al. (2000).}
\tiny
\begin{tabular}{lccrcrccccl}
\hline 
Name&Position&$\mu$&P.A&$\sigma_{\mu}$&$\sigma_{PA}$&$Y$&$J$&$H$&$K$&note\\
&&''/yr&$^{\circ}$&''/yr&$^{\circ}$&&&&&\\
  \hline
ULAS2MASS1017+0711&10 17 23.35 +07 11 03.9&0.159&158.0&0.024$^1$&6.7&15.396&14.656&14.103&13.695&f\\ 
ULAS2MASS1017+0719&10 17 26.72 +07 19 25.4&0.188&239.3&0.013$^2$&3.7&13.207&12.399&11.859&11.415&b\\ 
ULAS2MASS1029+0006&10 29 57.01 +00 06 44.6&0.117&281.6&0.014$^1$&8.3&15.947&15.170&14.648&14.215&\\ 
ULAS2MASS1126-0018&11 26 30.34 $-$00 18 32.0&0.088&343.5&0.013$^1$&14.2&15.386&14.639&14.048&13.642&\\ 
ULAS2MASS1129+0002&11 29 09.04 +00 02 56.8&0.159&291.6&0.019$^1$&5.1&15.049&14.295&13.731&13.291&\\ 
ULAS2MASS1134+0022&11 34 55.13 +00 22 51.9&0.511&129.1&0.014$^2$&1.5&13.668&12.774&12.171&11.684&h\\ 
ULAS2MASS1140+0007&11 40 06.24 +00 07 02.2&0.163&258.0&0.011$^2$&3.8&15.096&14.393&13.845&13.464&f\\ 
ULAS2MASS1147+0010&11 47 43.71 +00 10 28.1&0.130&235.5&0.017$^1$&7.7&16.096&15.370&14.828&14.433&\\ 
ULAS2MASS1151-0015&11 51 05.74 $-$00 15 29.7&0.289&190.8&0.011$^1$&3.5&15.466&14.711&14.211&13.762&\\ 
ULAS2MASS1159+0706&11 59 48.15 +07 06 59.1&0.201&301.1&0.023$^1$&6.2&16.027&15.304&14.801&14.399&\\ 
ULAS2MASS1201+0735&12 01 59.64 +07 35 53.7&0.164&135.8&0.026$^1$&8.5&15.864&15.025&14.428&13.946&\\ 
ULAS2MASS1207+0354&12 07 43.99 +03 54 30.8&0.157&139.7&0.012$^2$&4.3&14.727&13.992&13.409&13.004&\\ 
ULAS2MASS1208+0845&12 08 16.83 +08 45 27.6&0.132&251.0&0.022$^1$&5.6&14.796&13.958&13.350&12.920&\\ 
ULAS2MASS1211+0406&12 11 30.11 +04 06 08.1&0.225&195.7&0.035$^1$&6.5&16.808&15.517&14.680&13.949&\\ 
ULAS2MASS1218+1140&12 18 30.23 +11 40 14.4&0.422&170.6&0.011$^2$&1.6&14.824&14.090&13.587&13.184&\\ 
ULAS2MASS1223+0445&12 23 23.53 +04 45 36.3&0.097&174.0&0.012$^2$&8.7&13.967&13.256&12.709&12.320&\\ 
ULAS2MASS1224+1047&12 24 41.23 +10 47 45.0&0.089&293.0&0.014$^2$&7.7&14.188&13.487&12.986&12.603&\\ 
ULAS2MASS1225+1340&12 25 56.50 +13 40 03.9&0.115&243.6&0.011$^2$&5.4&14.940&14.104&13.520&13.062&\\ 
ULAS2MASS1226+1239&12 26 59.31 +12 39 06.1&0.122&138.5&0.021$^1$&9.8&16.159&15.261&14.683&14.227&\\ 
ULAS2MASS1227+0005&12 27 48.30 +00 05 43.7&0.197&222.8&0.033$^1$&9.6&15.876&15.067&14.535&14.069&i\\ 
ULAS2MASS1228+1039&12 28 29.03 +10 39 10.2&0.180&167.4&0.028$^1$&7.1&16.153&15.282&14.629&14.140&\\ 
ULAS2MASS1230+1340&12 30 19.20 +13 40 40.5&0.232&152.7&0.022$^1$&6.3&15.836&15.019&14.544&14.086&\\ 
ULAS2MASS1230+0559&12 30 51.83 +05 59 14.8&0.142&287.6&0.027$^1$&10.5&15.663&14.859&14.354&13.927&\\ 
ULAS2MASS1231+0847&12 31 46.99 +08 47 25.8&1.590&228.1&0.023$^1$&0.9&16.340&15.153&15.456&15.552&j\\ 
ULAS2MASS1232+1033&12 32 36.04 +10 33 00.2&0.129&174.8&0.018$^1$&4.0&14.065&13.323&12.773&12.366&\\ 
ULAS2MASS1234+0432&12 34 42.51 +04 32 44.9&0.123&281.8&0.019$^1$&9.5&15.455&14.637&14.092&13.679&\\ 
ULAS2MASS1235+0732&12 35 10.04 +07 32 12.9&0.101&251.6&0.012$^2$&6.6&15.626&14.914&14.394&14.000&\\ 
ULAS2MASS1244+0708&12 44 57.23 +07 08 23.7&0.350&146.2&0.018$^1$&3.7&15.821&15.049&14.637&14.125&\\ 
ULAS2MASS1245+0736&12 45 56.63 +07 36 54.6&0.184&152.6&0.011$^2$&3.6&14.725&14.010&13.458&13.038&\\ 
ULAS2MASS1246+1202&12 46 01.91 +12 02 43.2&0.220&240.6&0.012$^2$&3.0&15.761&14.908&14.335&13.864&\\ 
ULAS2MASS1246+1503&12 46 04.20 +15 03 40.3&0.078&181.4&0.008$^1$&14.7&15.717&14.994&14.445&14.057&\\ 
ULAS2MASS1253+0740&12 53 49.36 +07 40 04.5&0.162&269.6&0.026$^1$&5.6&14.559&13.807&13.323&12.870&\\ 
ULAS2MASS1253+0740&12 53 49.69 +07 40 00.6&0.156&269.5&0.029$^1$&7.6&14.977&14.168&13.669&13.181&\\ 
ULAS2MASS1257+0600&12 57 48.11 +06 00 30.3&0.172&269.8&0.031$^1$&9.8&16.429&15.650&15.073&14.678&\\ 
ULAS2MASS1258+0541&12 58 52.78 +05 41 19.9&0.139&323.7&0.022$^1$&10.3&14.870&14.138&13.578&13.143&\\ 
ULAS2MASS1259+0651&12 59 37.59 +06 51 18.9&0.444&240.3&0.032$^1$&5.0&15.045&14.289&13.851&13.422&\\ 
ULAS2MASS1259+1136&12 59 59.54 +11 36 18.3&0.088&285.0&0.012$^2$&7.4&15.644&14.902&14.330&13.895&\\ 
ULAS2MASS1303+1156&13 03 05.44 +11 56 17.7&0.129&208.2&0.015$^1$&5.9&15.653&14.770&14.150&13.680&\\ 
ULAS2MASS1308+0818&13 08 30.97 +08 18 52.5&0.233&281.1&0.022$^1$&3.4&16.312&15.192&14.373&13.792&\\ 
ULAS2MASS1310+1000&13 10 13.67 +10 00 20.3&0.121&252.9&0.019$^1$&15.4&16.362&15.631&15.028&14.638&\\ 
ULAS2MASS1313+0933&13 13 30.46 +09 33 58.8&0.129&271.2&0.016$^2$&5.7&14.127&13.426&12.875&12.494&\\ 
ULAS2MASS1320+0957&13 20 41.49 +09 57 49.7&0.256&238.7&0.037$^1$&6.8&14.484&13.654&13.111&12.627&n\\ 
ULAS2MASS1325+1011&13 25 50.38 +10 11 02.9&0.111&137.2&0.019$^1$&9.7&15.344&14.477&13.924&13.446&\\ 
ULAS2MASS1327+0916&13 27 26.77 +09 16 05.6&0.146&238.3&0.012$^1$&5.6&15.300&14.539&13.994&13.578&\\ 
ULAS2MASS1328+0758&13 28 43.25 +07 58 37.7&0.157&258.3&0.013$^1$&4.0&15.551&14.722&14.114&13.655&\\ 
ULAS2MASS1329+1116&13 29 08.65 +11 16 34.8&0.142&289.1&0.027$^1$&13.5&16.429&15.673&15.124&14.698&\\ 
ULAS2MASS1332+1019&13 32 35.33 +10 19 18.6&0.221&304.1&0.033$^1$&6.6&15.582&14.792&14.251&13.826&\\ 
ULAS2MASS1334+1221&13 34 37.24 +12 21 57.2&0.126&277.7&0.012$^2$&4.9&15.301&14.585&13.998&13.621&\\ 
ULAS2MASS1345+0849&13 45 10.68 +08 49 54.0&0.118&169.4&0.016$^1$&5.8&15.365&14.639&14.119&13.707&\\ 
ULAS2MASS1345+1150&13 45 15.73 +11 50 51.6&0.330&302.0&0.013$^1$&2.3&15.141&14.420&13.887&13.491&\\ 
ULAS2MASS1345+1134&13 45 16.48 +11 34 11.0&0.242&115.0&0.025$^1$&5.2&15.359&14.573&14.041&13.597&\\ 
ULAS2MASS1346+0842&13 46 07.37 +08 42 34.1&0.235&246.6&0.042$^1$&8.9&16.744&15.518&14.752&14.115&\\ 
ULAS2MASS1346-0006&13 46 40.44 $-$00 06 54.2&0.413&256.5&0.020$^2$&2.5&14.591&13.830&13.306&12.814&o\\ 
ULAS2MASS1352+1116&13 52 12.16 +11 16 13.5&0.230&239.6&0.036$^1$&11.4&15.228&14.360&13.745&13.273&\\ 
ULAS2MASS1352+0916&13 52 57.86 +09 16 47.3&0.263&162.0&0.015$^1$&2.1&15.787&14.799&14.142&13.553&\\ 
ULAS2MASS1354+0846&13 54 08.67 +08 46 08.7&0.230&283.7&0.014$^2$&3.2&12.976&12.146&11.587&11.144&f\\ 
ULAS2MASS1404+0740&14 04 09.65 +07 40 08.4&0.546&188.3&0.015$^2$&1.7&13.496&12.783&12.256&11.881&k\\ 
ULAS2MASS1407+0837&14 07 18.26 +08 37 59.7&0.151&228.7&0.024$^1$&9.9&14.827&14.102&13.527&13.153&\\ 
ULAS2MASS1407+1241&14 07 53.45 +12 41 10.4&0.337&280.5&0.022$^1$&3.3&16.737&15.333&14.392&13.632&n\\ 
ULAS2MASS1418+0808&14 18 44.33 +08 08 46.7&0.139&270.7&0.018$^1$&10.4&15.744&15.032&14.468&14.086&\\ 
ULAS2MASS1422+0827&14 22 57.10 +08 27 50.4&0.592&195.6&0.012$^1$&1.5&16.240&15.009&14.271&13.609&\\ 
ULAS2MASS1425+0712&14 25 53.72 +07 12 15.4&0.203&200.2&0.026$^1$&4.5&15.683&14.959&14.518&14.119&\\ 
ULAS2MASS1429+0718&14 29 29.38 +07 18 42.1&0.203&255.1&0.031$^1$&11.4&16.224&15.255&14.687&14.283&\\ 
ULAS2MASS1432+1312&14 32 46.67 +13 12 01.5&0.153&272.8&0.019$^1$&8.7&16.288&15.491&14.915&14.487&\\ 
ULAS2MASS1433+1334&14 33 43.64 +13 34 46.3&0.088&159.1&0.015$^1$&15.3&15.841&15.139&14.598&14.231&\\ 
ULAS2MASS1436+0920&14 36 59.10 +09 20 52.9&0.229&149.3&0.040$^1$&10.3&15.190&14.487&13.967&13.565&f\\ 
ULAS2MASS1440+1252&14 40 07.51 +12 52 18.1&0.303&271.6&0.017$^1$&2.4&15.568&14.815&14.269&13.841&\\ 
ULAS2MASS1440+1233&14 40 30.22 +12 33 33.3&0.174&225.3&0.016$^1$&5.0&15.309&14.362&13.687&13.134&o\\ 
ULAS2MASS1441+1244&14 41 08.87 +12 44 24.9&0.147&283.2&0.021$^1$&8.5&16.287&15.454&14.866&14.420&\\ 
ULAS2MASS1442+1239&14 42 32.82 +12 39 26.5&0.154&297.5&0.017$^1$&7.8&16.106&15.377&14.835&14.433&\\ 
ULAS2MASS1448+1031&14 48 25.75 +10 31 58.1&0.278&117.5&0.018$^1$&5.1&15.804&14.420&13.510&12.674&m\\ 
ULAS2MASS1448+1048&14 48 34.11 +10 48 05.0&0.133&179.4&0.013$^2$&6.2&13.582&12.856&12.371&11.974&\\ 
ULAS2MASS1451+0644&14 51 32.00 +06 44 53.0&0.150&182.9&0.026$^1$&6.9&16.076&15.137&14.501&13.963&\\ 
ULAS2MASS1452+0931&14 52 01.34 +09 31 35.9&0.201&161.3&0.037$^1$&9.3&16.346&15.337&14.809&14.279&\\ 
ULAS2MASS1452+1114&14 52 01.96 +11 14 56.9&0.394&139.4&0.027$^1$&4.1&16.771&15.569&14.867&14.280&\\ 
ULAS2MASS1452+0723&14 52 15.47 +07 23 52.0&0.290&236.9&0.017$^1$&3.4&15.024&14.317&13.840&13.463&\\ 
ULAS2MASS1457+1102&14 57 30.56 +11 02 01.8&0.117&281.9&0.011$^1$&13.1&13.169&12.466&12.013&11.616&\\ 
ULAS2MASS1501-0236&15 01 57.05 $-$02 36 06.9&0.243&180.9&0.018$^1$&5.2&15.841&15.126&14.586&14.183&c\\ 
 \hline
\end{tabular}
\end{table*}
\addtocounter{table}{-1}
\begin{table*}
   \caption{\label{LASobjects} Objects identified in our LAS sample. All photometry is in the UKIDSS system (Hewett et al. 2006). $^1$ denotes objects whose astrometric solutions was calculated using local reference stars while $^2$ denotes those calculated using only global error estimates. Citation key - a:Schneider et al. (2002) b: Luyten (1979b) c: Tinney (1993) d: Fan et al. (2000) e: Bouy et al. (2003) f: Lepine \& Shara (2005) g: Cruz et al. (2003) h: Hawley et al. (2002) i: Delfosse et al. (1999) j: Burgasser et al. (2004) k: Luyten (1979a) l Cruz et al. (2007) m: Wilson et al. (2003) n: Reid et al. (2008), p: Knapp et al. (2004), q: Gizis et al. (2000), r:Kirkpatrick et al. (2000).}
\tiny
\begin{tabular}{lccrcrccccl}
\hline 
Name&Position&$\mu$&P.A&$\sigma_{\mu}$&$\sigma_{PA}$&$Y$&$J$&$H$&$K$&note\\
&&''/yr&$^{\circ}$&''/yr&$^{\circ}$&&&&&\\
  \hline
ULAS2MASS1504+0923&15 04 10.17 +09 23 23.9&0.271&289.7&0.028$^1$&6.7&13.738&13.028&12.518&12.077&b\\ 
ULAS2MASS1505+1029&15 05 56.42 +10 29 40.0&0.106&183.4&0.016$^1$&13.5&15.387&14.682&14.119&13.707&\\ 
ULAS2MASS1510-0241&15 10 16.67 $-$02 41 07.7&0.399&272.2&0.014$^2$&1.8&13.494&12.543&11.981&11.315&c\\ 
ULAS2MASS1511-0215&15 11 53.59 $-$02 15 31.8&0.145&105.8&0.017$^1$&8.2&15.629&14.923&14.414&13.986&\\ 
ULAS2MASS1513+0844&15 13 58.26 +08 44 34.4&0.136&157.1&0.018$^1$&7.9&15.624&14.753&14.186&13.762&\\ 
ULAS2MASS1514+1201&15 14 13.77 +12 01 45.0&0.161&260.4&0.018$^1$&6.6&15.795&14.803&14.230&13.732&\\ 
ULAS2MASS1516+0942&15 16 39.94 +09 42 11.6&0.181&286.7&0.020$^1$&4.7&14.626&13.696&13.170&12.689&\\ 
ULAS2MASS1518+0849&15 18 35.32 +08 49 08.0&0.187&139.2&0.027$^1$&8.1&16.157&15.364&14.905&14.434&\\ 
ULAS2MASS1518+0807&15 18 49.12 +08 07 43.5&0.175&216.5&0.021$^1$&6.5&15.831&15.129&14.649&14.247&\\ 
ULAS2MASS1520+0407&15 20 46.08 +04 07 52.3&0.097&195.6&0.013$^1$&10.2&14.418&13.635&13.070&12.676&\\ 
ULAS2MASS1521+0653&15 21 05.19 +06 53 07.0&0.120&277.8&0.022$^1$&9.7&15.955&15.177&14.569&14.154&\\ 
ULAS2MASS1524+1019&15 24 26.65 +10 19 49.4&0.114&308.5&0.016$^1$&8.1&14.771&14.015&13.473&13.027&\\ 
ULAS2MASS1524+0934&15 24 32.04 +09 34 37.6&0.155&169.0&0.022$^1$&5.5&15.762&14.948&14.442&14.048&\\ 
ULAS2MASS1526+0745&15 26 54.56 +07 45 41.2&0.174&206.0&0.021$^1$&6.2&16.006&15.139&14.496&14.022&\\ 
ULAS2MASS1529+0637&15 29 58.14 +06 37 00.3&0.194&225.7&0.030$^1$&8.9&16.212&15.500&15.046&14.629&\\ 
ULAS2MASS1531+0644&15 31 46.98 +06 44 23.1&0.164&123.2&0.023$^1$&7.4&14.847&14.142&13.604&13.218&f\\ 
ULAS2MASS1531+0915&15 31 54.73 +09 15 59.9&0.128&256.5&0.023$^1$&11.0&16.407&15.693&15.192&14.772&\\ 
ULAS2MASS1532+0837&15 32 17.20 +08 37 12.5&0.175&184.6&0.019$^1$&6.3&16.208&15.368&14.730&14.250&\\ 
ULAS2MASS1535+0932&15 35 41.48 +09 32 02.4&0.109&259.7&0.018$^1$&22.4&16.235&15.414&14.896&14.451&\\ 
ULAS2MASS1537+0604&15 37 05.80 +06 04 04.2&0.148&199.5&0.022$^1$&8.1&15.101&14.268&13.650&13.183&\\ 
ULAS2MASS1541+0620&15 41 43.77 +06 20 43.7&0.141&261.1&0.020$^1$&10.5&16.684&15.733&15.188&14.712&\\ 
ULAS2MASS1543+0939&15 43 15.91 +09 39 35.5&0.165&261.8&0.013$^1$&5.7&15.665&14.939&14.406&13.971&\\ 
ULAS2MASS1545-0124&15 45 13.52 $-$01 24 48.6&0.218&102.8&0.015$^1$&3.4&14.858&14.032&13.540&13.156&\\ 
ULAS2MASS1546+0317&15 46 25.93 +03 17 54.8&0.113&277.7&0.021$^1$&11.9&13.951&13.144&12.563&12.096&n\\ 
ULAS2MASS1547+1023&15 47 43.51 +10 23 31.8&0.156&232.4&0.015$^1$&6.0&15.710&14.971&14.477&14.044&\\ 
ULAS2MASS1553+0749&15 53 08.58 +07 49 30.6&0.123&303.0&0.021$^1$&9.7&14.908&14.054&13.491&13.050&\\ 
ULAS2MASS2143-0040&21 43 41.24 $-$00 40 44.4&0.123&54.5&0.017$^1$&10.5&14.598&13.883&13.369&12.998&\\ 
ULAS2MASS2147-0029&21 47 21.03 $-$00 29 46.3&0.186&26.2&0.027$^1$&10.4&15.274&14.535&13.942&13.529&\\ 
ULAS2MASS2148+0020&21 48 30.90 +00 20 53.8&0.177&140.8&0.020$^1$&6.7&16.314&15.397&14.748&14.244&\\ 
ULAS2MASS2203+0024&22 03 23.83 +00 24 05.5&0.163&235.8&0.012$^1$&4.5&15.738&14.979&14.467&14.051&c\\ 
ULAS2MASS2206-0037&22 06 09.69 $-$00 37 27.9&0.188&186.2&0.009$^1$&4.2&15.254&14.542&14.015&13.616&c\\ 
ULAS2MASS2208-0003&22 08 17.74 $-$00 03 09.2&0.126&89.0&0.022$^1$&7.6&15.399&14.695&14.122&13.742&\\ 
ULAS2MASS2214+0052&22 14 59.79 +00 52 34.4&0.241&96.1&0.021$^1$&3.0&14.731&14.000&13.479&13.106&c\\ 
ULAS2MASS2221-0022&22 21 16.88 $-$00 22 16.8&0.128&116.0&0.022$^1$&13.7&15.687&14.854&14.256&13.823&c\\ 
ULAS2MASS2237-0039&22 37 39.94 $-$00 39 50.8&0.122&129.8&0.017$^1$&7.7&14.713&13.913&13.357&12.933&c\\ 
ULAS2MASS2249+0025&22 49 35.39 +00 25 58.2&0.187&116.5&0.016$^1$&4.8&15.221&14.513&13.908&13.532&c\\ 
ULAS2MASS2258+0113&22 58 54.07 +01 13 49.9&0.252&164.3&0.013$^1$&3.2&14.695&13.875&13.365&12.883&\\ 
ULAS2MASS2300+0103&23 00 52.47 +01 03 13.8&0.271&121.1&0.028$^1$&5.0&15.704&14.838&14.228&13.733&\\ 
ULAS2MASS2304+0749&23 04 25.97 +07 49 00.9&0.237&96.7&0.022$^1$&5.0&15.414&14.464&13.817&13.330&\\ 
ULAS2MASS2308+0035&23 08 39.85 +00 35 34.3&0.218&123.0&0.035$^1$&10.0&16.399&15.550&14.939&14.497&\\ 
ULAS2MASS2314+0623&23 14 09.89 +06 23 09.4&0.161&90.7&0.031$^1$&10.4&16.428&15.613&15.027&14.542&\\ 
ULAS2MASS2316+1339&23 16 37.93 +13 39 15.3&0.091&241.3&0.013$^1$&8.1&15.895&15.185&14.590&14.179&\\ 
ULAS2MASS2318+1305&23 18 18.42 +13 05 03.8&0.142&217.6&0.016$^1$&6.6&15.212&14.471&13.937&13.560&\\ 
ULAS2MASS2321+0815&23 21 30.10 +08 15 45.2&0.137&88.8&0.025$^1$&11.4&16.037&15.249&14.706&14.289&\\ 
ULAS2MASS2331-0005&23 31 01.72 $-$00 05 30.1&0.251&98.3&0.012$^1$&4.3&15.728&14.998&14.496&14.082&\\ 
ULAS2MASS2331+1552&23 31 29.29 +15 52 21.9&0.158&234.8&0.012$^1$&4.1&16.034&14.976&14.303&13.736&\\ 
ULAS2MASS2332-0050&23 32 24.41 $-$00 50 25.2&0.098&86.5&0.013$^2$&6.6&14.330&13.555&13.013&12.579&\\ 
ULAS2MASS2333+0050&23 33 58.48 +00 50 12.2&0.168&76.7&0.015$^1$&4.6&15.879&14.934&14.359&13.851&\\ 
ULAS2MASS2338+1604&23 38 24.77 +16 04 58.6&0.193&31.9&0.016$^1$&5.2&15.677&14.936&14.402&13.978&\\ 
ULAS2MASS2345+0055&23 45 39.08 +00 55 13.4&0.115&112.1&0.014$^2$&6.3&14.632&13.677&13.087&12.523&n\\ 
ULAS2MASS2350+1441&23 50 49.62 +14 41 37.2&0.107&56.4&0.012$^1$&7.4&15.137&14.381&13.809&13.383&\\ 
ULAS2MASS2353+1511&23 53 38.48 +15 11 49.4&0.260&102.6&0.019$^1$&3.4&15.434&14.680&14.137&13.717&\\ 
ULAS2MASS2355+0115&23 55 42.54 +01 15 22.2&0.173&61.9&0.028$^1$&8.1&15.025&14.251&13.723&13.374&\\ 
ULAS2MASS2355+0016&23 55 46.28 +00 16 27.9&0.136&180.5&0.024$^1$&7.6&15.045&14.276&13.746&13.344&\\ 
ULAS2MASS2356-0034&23 56 17.60 $-$00 34 33.8&0.120&216.8&0.019$^1$&9.6&16.188&15.473&14.927&14.520&\\ 
ULAS2MASS2356-0058&23 56 34.70 $-$00 58 16.3&0.142&223.6&0.023$^1$&9.3&16.468&15.710&15.204&14.825&\\ 
\hline
\end{tabular}
\end{table*}
\begin{table*}
   \caption{\label{GCSobjects} Objects identified in our GCS sample. All photometry is in the UKIDSS system (Hewett et al. 2006). $^1$ denotes objects whose astrometric solutions was calculated using local reference stars while $^2$ denotes those calculated using only global error estimates. $^*$ denotes the object has been identified as a potential cluster member. a: Stauffer et al. (1999) b: Barrado et al. (2002) c: Cossburn et al. (1997) d: Kirkpatrick et al. (1997) e: Ambartsmunyan et al. (1973) f: Bouvier et al. (1998) g: Lepine \& Shara (2005) h: Kirkpatrick et al. (2000) i: Luyten (1979b) j: Lepine, Shara \& Rich (2002)}
\tiny
\begin{tabular}{lccrcrccccl}
\hline
Name&Position&$\mu$&P.A&$\sigma_{\mu}$&$\sigma_{PA}$&$Y$&$J$&$H$&$K$&note\\
&&''/yr&$^{\circ}$&''/yr&$^{\circ}$&&&&&\\
  \hline
UGCS2MASS0319+5030&03 19 41.42 +50 30 43.8&0.194&147.1&0.008$^1$&2.3&16.109&15.210&14.532&14.013&a*\\
UGCS2MASS0319+4848&03 19 51.84 +48 48 22.0&0.220&247.6&0.008$^1$&2.2&14.748&13.899&13.368&12.906&b*\\
UGCS2MASS0328+4841&03 28 02.35 +48 41 05.8&0.309&130.7&0.013$^1$&2.4&15.629&14.896&14.415&13.980&b*\\
UGCS2MASS0335+4809&03 35 38.61 +48 09 20.7&0.110&142.0&0.014$^1$&7.4&14.569&13.829&13.263&12.836&\\
UGCS2MASS0336+2233&03 36 25.95 +22 33 17.2&0.243&132.7&0.007$^1$&1.7&15.808&14.991&14.471&13.977&\\
UGCS2MASS0339+2457&03 39 52.92 +24 57 27.0&0.178&107.6&0.028$^1$&4.2&13.622&12.786&12.212&11.744&d\\
UGCS2MASS0340+2215&03 40 43.38 +22 15 07.0&0.154&268.4&0.014$^2$&4.1&13.819&13.089&12.545&12.142&e*\\
UGCS2MASS0341+3207&03 41 16.45 +32 07 55.2&0.137&214.2&0.008$^1$&3.6&15.905&15.137&14.594&14.093&\\
UGCS2MASS0342+2248&03 42 10.17 +22 48 44.6&0.089&166.9&0.015$^1$&10.7&16.355&15.543&14.932&14.469&\\
UGCS2MASS0343+3037&03 43 57.35 +30 37 19.0&0.083&155.5&0.006$^1$&6.5&15.612&14.763&14.117&13.586&\\
UGCS2MASS0345+2540&03 45 43.12 +25 40 23.1&0.109&248.5&0.020$^1$&6.0&15.009&13.924&13.240&12.663&c*\\
UGCS2MASS0347+3406&03 47 54.07 +34 06 47.7&0.096&153.7&0.010$^1$&5.9&16.262&15.304&14.701&14.139&\\
UGCS2MASS0350+3149&03 50 02.97 +31 49 58.8&0.124&62.5&0.009$^1$&4.1&16.255&15.490&14.972&14.516&\\
UGCS2MASS0351+2544&03 51 57.26 +25 44 15.0&0.373&109.0&0.024$^1$&3.3&15.045&14.343&13.928&13.504&\\
UGCS2MASS0352+2359&03 52 07.88 +23 59 13.1&0.091&165.4&0.016$^1$&5.7&15.371&14.636&14.066&13.640&f*\\
UGCS2MASS0354+2316&03 54 01.47 +23 16 33.5&0.206&107.0&0.009$^1$&5.1&13.895&13.046&12.452&11.980&d*\\
UGCS2MASS0401+2322&04 01 43.07 +23 22 12.4&0.112&126.8&0.010$^1$&4.8&14.151&13.413&12.869&12.442&\\
UGCS2MASS0402+2837&04 02 58.77 +28 37 53.5&0.099&115.2&0.013$^1$&5.3&15.789&15.069&14.549&14.114&\\
UGCS2MASS0403+2616&04 03 07.86 +26 16 08.0&0.654&135.4&0.007$^1$&0.7&14.207&13.489&13.068&12.708&j\\
UGCS2MASS0407+2734&04 07 48.22 +27 34 05.2&0.088&191.3&0.014$^1$&11.9&16.183&15.453&14.918&14.509&\\
UGCS2MASS0408+2447&04 08 40.01 +24 47 32.3&0.081&139.1&0.008$^1$&5.3&16.275&15.499&14.592&14.199&\\
UGCS2MASS0409+2104&04 09 09.57 +21 04 37.9&0.184&150.6&0.007$^1$&2.4&16.622&15.376&14.502&13.781&h\\
UGCS2MASS0409+2446&04 09 33.19 +24 46 33.7&0.144&146.3&0.011$^1$&4.5&16.268&15.468&14.878&14.376&\\
UGCS2MASS0427+2900&04 27 23.86 +29 00 27.0&0.102&142.3&0.011$^1$&6.5&16.296&15.520&14.862&14.448&\\
UGCS2MASS0433+2933&04 33 37.21 +29 33 13.1&0.120&180.6&0.010$^1$&2.7&14.395&13.608&12.651&12.317&\\
UGCS2MASS0434+2937&04 34 54.42 +29 37 22.2&0.239&137.8&0.009$^1$&2.2&16.325&15.595&14.911&14.522&\\
UGCS2MASS0435+2956&04 35 23.50 +29 56 27.4&0.084&154.1&0.010$^1$&5.5&15.415&14.674&13.942&13.522&\\
UGCS2MASS0436+3019&04 36 26.30 +30 19 56.3&0.115&129.5&0.010$^1$&5.1&16.030&15.312&14.840&14.433&\\
UGCS2MASS0535-0551&05 35 06.67 $-$05 51 09.8&0.159&173.7&0.016$^1$&5.8&15.599&14.884&14.344&13.920&\\
UGCS2MASS0536-0454&05 36 16.38 $-$04 54 34.2&0.083&101.3&0.013$^1$&9.8&16.418&15.623&14.714&14.318&\\
UGCS2MASS0536-0328&05 36 37.18 $-$03 28 50.8&0.354&157.0&0.017$^1$&2.2&15.053&14.186&13.554&13.087&\\
UGCS2MASS0537-0403&05 37 47.55 $-$04 03 26.9&0.157&184.2&0.012$^1$&5.9&15.300&14.564&13.972&13.564&\\
UGCS2MASS0541-0305&05 41 55.81 $-$03 05 14.2&0.168&324.2&0.009$^1$&3.1&16.387&15.381&14.271&13.784&\\
UGCS2MASS0830+1821&08 30 29.92 +18 21 22.8&0.118&124.4&0.018$^1$&7.7&15.751&14.821&14.185&13.693&\\
UGCS2MASS0831+2036&08 31 28.42 +20 36 54.1&0.097&177.5&0.010$^1$&6.9&16.136&15.365&14.900&14.477&\\
UGCS2MASS0835+2030&08 35 29.18 +20 30 49.0&0.111&193.8&0.018$^1$&7.0&14.555&13.829&13.262&12.871&\\
UGCS2MASS0835+2224&08 35 45.28 +22 24 30.9&0.159&261.5&0.015$^1$&5.8&16.490&15.559&14.927&14.443&\\
UGCS2MASS0838+1852&08 38 08.04 +18 52 04.9&0.107&184.1&0.017$^1$&7.5&15.620&14.814&14.259&13.844&\\
UGCS2MASS0840+2024&08 40 48.29 +20 24 10.4&0.411&121.8&0.012$^1$&2.0&16.411&15.661&15.248&14.871&\\
UGCS2MASS0848+2150&08 48 16.74 +21 50 51.0&0.339&152.1&0.019$^1$&3.3&14.458&13.738&13.229&12.850&g\\
UGCS2MASS1548-2148&15 48 45.71 $-$21 48 08.8&0.156&199.7&0.017$^1$&7.4&16.491&15.786&15.256&14.849&\\
UGCS2MASS1550-2201&15 50 11.50 $-$22 01 21.9&0.082&194.1&0.016$^1$&13.9&16.337&15.548&14.980&14.573&\\
UGCS2MASS1620-2847&16 20 03.77 $-$28 47 21.5&0.100&221.6&0.010$^1$&5.8&14.852&14.088&13.553&13.130&\\
UGCS2MASS1625-2117&16 25 46.88 $-$21 17 26.3&0.112&164.6&0.017$^1$&7.4&16.253&15.507&14.888&14.431&\\
UGCS2MASS1625-2400&16 25 50.24 $-$24 00 08.5&0.178&261.1&0.016$^2$&3.6&12.799&11.892&10.973&10.545&i\\
UGCS2MASS1626-2352&16 26 25.13 $-$23 52 37.8&0.133&100.4&0.021$^2$&6.5&12.585&11.802&11.072&10.704&\\
UGCS2MASS1628-2116&16 28 56.61 $-$21 16 09.9&0.097&152.5&0.017$^1$&9.1&14.650&13.935&13.168&12.747&\\
UGCS2MASS1630-2120&16 30 17.69 $-$21 20 01.4&0.153&259.8&0.013$^1$&3.8&15.695&14.524&13.780&13.174&\\
UGCS2MASS1631-2404&16 31 46.42 $-$24 04 47.5&0.104&87.5&0.013$^1$&5.9&16.530&15.735&14.859&14.396&\\
UGCS2MASS1632-2358&16 32 21.35 $-$23 58 23.8&0.160&197.2&0.012$^1$&4.8&16.390&15.554&14.993&14.494&\\
UGCS2MASS1637-2200&16 37 53.00 $-$22 00 14.7&0.084&252.1&0.011$^1$&9.4&15.997&15.112&14.259&13.821&\\
UGCS2MASS1639-2502&16 39 29.17 $-$25 02 12.3&0.093&142.9&0.014$^1$&7.5&16.189&15.466&14.800&14.489&\\
UGCS2MASS1640-2442&16 40 22.44 $-$24 42 48.5&0.083&209.8&0.014$^1$&10.8&16.425&15.599&14.930&14.441&\\
UGCS2MASS1751+0504&17 51 00.24 +05 04 23.0&0.173&204.4&0.029$^1$&9.1&16.752&15.733&15.073&14.503&\\
\hline
\\
\end{tabular}
\end{table*}

\normalsize
\label{lastpage}
\end{document}